\documentclass[11pt]{article}
\pdfoutput=1

\usepackage{cite}
\usepackage{booktabs}
\usepackage[english]{babel}
\usepackage{amsmath,amssymb,amsbsy,amstext, amsthm, simplewick}
\usepackage{hyperref}
\usepackage{graphicx}
\usepackage{amsfonts}
\usepackage{amssymb}
\usepackage[small]{caption}
\usepackage{upgreek}
\usepackage[svgnames,dvipsnames,x11names,table]{xcolor}
\usepackage{multirow}
\usepackage{geometry}
\usepackage[hang,flushmargin]{footmisc}
\usepackage{bm}
\usepackage{braket}
\usepackage{subcaption}
\usepackage{mathtools}
\usepackage{setspace}
\usepackage{cleveref}
\usepackage{comment}
\usepackage{scalerel}
\usepackage[normalem]{ulem}
\usepackage{slashed}
\usepackage{enumitem}
\usepackage{cancel}
\usepackage{dsfont}
\usepackage{tikz}
\usepackage{xcolor}
\usetikzlibrary{decorations.markings}
\usetikzlibrary{shapes.misc}

\makeatletter
\g@addto@macro\bfseries{\boldmath}
\makeatother

\hypersetup{
    colorlinks=true,
    linkcolor={red!50!black},
    citecolor={blue!50!black},
    urlcolor={blue!80!black}
}

\usepackage{colortbl}

\makeatletter
\newlength{\apb@width}
\newcommand{\autoparbox}[2][c]{\settowidth{\apb@width}{#2}\parbox[#1]{\apb@width}{#2}}

\makeatother

\definecolor{lightgray}{gray}{0.9}

\usepackage[framemethod=default]{mdframed}
\newmdenv[skipabove=7pt,
skipbelow=7pt,
rightline=false,
leftline=false,
topline=false,
bottomline=false,
backgroundcolor=gray!10,
linecolor=gray,
innerleftmargin=4pt,
innerrightmargin=4pt,
innertopmargin=4pt,
innerbottommargin=4pt,
leftmargin=0cm,
rightmargin=0cm,
linewidth=4pt]{eBox}

\usepackage[most]{tcolorbox}
\tcbset{colback=white, colframe=black,
        highlight math style= {enhanced, 
            colframe=red,colback=red!10!white,boxsep=0pt}
        }
\definecolor{light-gray}{gray}{0.95}

\crefname{table}{Table}{Tables}
\crefname{equation}{Eq.}{Eqs.}
\crefname{appendix}{App.}{Apps.}
\crefname{section}{Sec.}{Secs.}
\crefname{figure}{Fig.}{Figs.}

\numberwithin{equation}{section}

\def\beq{\begin{equation}}
\def\eeq{\end{equation}}

\def\bea{\begin{eqnarray}}
\def\eea{\end{eqnarray}}

\def\vp{\varphi_{+}}

\def\Tr{{\rm Tr}}

\def\beq{\begin{equation}}
\def\eeq{\end{equation}}
\def\bea{\begin{eqnarray}}
\def\eea{\end{eqnarray}}

\def\Tr{{\rm Tr}}

\def\O{{\cal O}}

\def\vp{{\varphi}}

\def\k{{\vec{\scaleto{k}{7pt}}}}

\def\p{{\vec p}}

\def\x{{\vec x}}

\def\bv{{\bar \varphi}}

\DeclareRobustCommand{\SkipTocEntry}[4]{}

\newcommand{\vev}[1]{\left\langle #1 \right\rangle}

\definecolor{colorTC}{rgb}{.2,.7,.2}
\definecolor{amethyst}{rgb}{0.6, 0.4, 0.8}

\definecolor{acolor}{rgb}{0.4, 0.2, 0.4}

\definecolor{blue3}{RGB}{31, 119, 180}
\definecolor{red3}{RGB}{	214, 39, 40}
\definecolor{orange3}{RGB}{255, 127, 14}
\definecolor{green3}{RGB}{44, 160, 44}

\begin{document}

\begin{titlepage}
\setcounter{page}{1} \baselineskip=15.5pt
\thispagestyle{empty}
$\quad$
\vskip 70 pt

\begin{center}
{\fontsize{20.74}{18} \bf Quantum Walks and Exact RG \\[5pt] in de Sitter Space}
\end{center}

\vskip 20pt
\begin{center}
\noindent
{\fontsize{12}{18}\selectfont Daniel Green and Kshitij Gupta}
\end{center}

\begin{center}
\vskip 4pt
\textit{$^1${\small Department of Physics, University of California at San Diego,  La Jolla, CA 92093, USA}}

\end{center}

\vspace{0.4cm}
 \begin{center}{\bf Abstract}
 \end{center}

\noindent

The local physics of light scalar fields in de Sitter space is well described by classical random walks, as expressed through the framework of Stochastic Inflation. Recent work has clarified how this formalism arises from quantum field theory (QFT) and the renormalization group (RG), allowing for corrections to this formalism to be determined order by order. Yet, this description is incomplete. For example, the quantum dynamics of these fields are expected to become important when determining the tail of the probability distribution for the fluctuations. In this paper, we develop the understanding of fields in de Sitter as a quantum walk in order to bridge the gap between the classical and quantum description. We use the framework of exact RG to calculate the evolution equation for the reduced density matrix of the long wavelength fields. This master equation provides the direct map from light fields to models of quantum walks. We show how to reduce the master equation to Stochastic Inflation, and provide a new understanding of how the higher-order corrections arise. In the process, we demonstrate that divergences and secular growth in de Sitter, for both light and heavy fields, can be absorbed by (dynamical) renormalization.

\end{titlepage}
\setcounter{page}{2}

\restoregeometry

\begin{spacing}{1.2}
\newpage
\setcounter{tocdepth}{2}
\tableofcontents
\end{spacing}

\setstretch{1.1}
\newpage

\section{Introduction}

A complete quantum description of our Universe remains one of the key ambitions of theoretical physics~\cite{Flauger:2022hie}. It requires reckoning with some of the most basic conceptual questions about the origin of the Universe, the predictions of quantum gravity, and the interpretation of quantum mechanics in closed systems.

The dynamics of light scalar fields present a controlled opportunity to address some of these essential questions. This first challenge is that massless scalar fields do not have a de Sitter (dS) invariant vacuum state~\cite{Allen:1985ux}. In addition, perturbative corrections exhibit secular growth that causes the perturbative expansion to break down~\cite{Akhmedov:2013vka,Akhmedov:2019cfd}.  It was suggested long ago~\cite{Vilenkin:1983xq,Starobinsky:1986fx} that these features of fields in de Sitter should be interpreted as consequences of local field values undergoing a classical random walk. In this language, the absence of a de Sitter invariant state and the secular growth become familiar aspects of random walks without a potential~\cite{Starobinsky:1994bd}. 

Recently, it has been shown that the random walk description of light scalars can be recovered in QFT on a fixed dS background using RG techniques~\cite{Burgess:2015ajz,LopezNacir:2016gzi,Gorbenko:2019rza,Baumgart:2019clc,Mirbabayi:2019qtx,Cohen:2020php,Mirbabayi:2020vyt,Baumgart:2020oby,Palma:2023uwo,Palma:2025oux}. In the process, many of these long standing issues are resolved at both a technical and conceptual level. From the RG point of view, the theory evolves away from the trivial fixed point to a non-trivial self-similar description that is equivalent to a random walk. At the same time, this description must have a limited regime of validity; a classical random walk generates entropy which cannot be the case in a closed quantum system. The stochastic formalism is therefore limited to subsystems within de Sitter space~\cite{Chandrasekaran:2022cip,Chen:2024rpx,Kudler-Flam:2024psh}, although that aspect is not always transparent in the derivation. A complete understanding of light fields in de Sitter should furnish an explanation for the origin and breakdown of this description.

A more complete quantum stochastic description of the long modes in de Sitter is nearly as old~\cite{Habib:1990hx,LaFlamme:1990kd,Nambu:1991vs,Prokopec:1992ia,Polarski:1995jg} as the original, classical description. Rather than calculating a probability distribution in the field basis, one calculates the evolution for the reduced density matrix of the long wavelength fields. Tracing over the short wavelength modes gives rise to an evolution equation. When expressed in terms of the Wigner quasi-probability distribution, this evolution remains in a random walk form in ``phase space". This approach has been revived more recently in~\cite{Kiefer:2006je,Weenink:2011dd,Tajima:2024tzu,Li:2025azq,Cespedes:2025zqp}, motivated in part by recent progress understanding the quantum nature of de Sitter space.

Despite this long history, the quantum description of light fields remains under-developed. For one, the stochastic evolution equation appears even in trivial free field theories, where the short and long wavelength models are uncorrelated. In this regard, it cannot be the case that the random walk arises only from tracing over short wavelength modes. Instead, one is performing an additional trace, at least implicitly, to arrive at a Fokker-Planck equation. Although one recovers the expected evolution to leading order, the quantum mechanical nature of this evolution equation depends on this procedure. Many of these subtleties arise even in flat space, and are a consequence of working with EFT observables within the in-in formalism~\cite{Salcedo:2022aal, Green:2024cmx,Salcedo:2024smn,Cespedes:2025ple}. These challenges make it impossible for current techniques to determine the tail of the local one-point distribution~\cite{Cohen:2022clv,Calderon-Figueroa:2025dto}, or whether the universe is eternally inflating~\cite{Cohen:2021jbo}. These types of questions require a non-perturbative understanding of the quantum evolution of the density matrix.

In this paper, we will use the framework of quantum walks~\cite{Venegas-Andraca:2012zkr} to better understand the quantum nature of the long wavelength fluctuations in de Sitter. A walker can be coupled to  $N$-quantum coins (or $N$-harmonic oscillators) so that the location of the walker is entangled with the space of the coins. Tracing over the coins gives rise to different types of stochastic behavior, including both purely classical and uniquely quantum walks, depending on how the systems are coupled. The coins alone may be in a product state so that no entanglement entropy is generated until they are coupled to the walker. In fact, we will show that the origin of Stochastic Inflation is most easily derived by introducing a local physical measurement similar to the walker.

We will then use the framework of exact RG~\cite{Polchinski:1983gv} to show that light fields in de Sitter space obey the equations of a general quantum walk. We will show how integrating out a shell of momentum gives rise to an evolution equation for the reduced density matrix, generalizing previous results from flat space~\cite{Goldman:2024cvx,Green:2024cmx} (due in part to momentum space entanglement~\cite{Balasubramanian:2011wt,Grignani:2016igg,Green:2022fwg,Flynn:2022tbj,Costa:2022bvs}). This master equation for the evolution of the density matrix only gives rise to the physics of Stochastic Inflation after we sum over Fourier modes (tracing over information at separated points). After deriving an exact equation for the evolution, we will show that this equation can be approximated by the original equations of Stochastic Inflation and its generalization to the density matrix and Wigner distribution. The resulting evolution equations are most intuitive in the Wigner representation, where we find an evolution equation qualitatively similar to the Caldeira-Leggett model~\cite{PhysRevLett.46.211} and quantum Brownian motion~\cite{Grabert:1988yt,Hu:1991di}, but that differs in several key respects.

As a byproduct of our exact RG analysis, we demonstrate that the divergences in cosmology can be absorbed into the definition of the action and operators in precise analogy with RG in flat space. Unlike flat space, the RG transformation involves time evolution giving rise to dynamical RG flow. This confirms the intuition that large logs found in perturbation theory can, and should be, resummed using RG techniques~\cite{Burgess:2009bs,Green:2020txs}.

The organization of this paper is as follows: In Section~\ref{sec:walks}, we discuss simple models of quantum walks that capture both conventional classical random walks, and uniquely quantum contributions. In Section~\ref{sec:exact}, we derive the exact RG equations for light fields in dS. We then show how they combine with time evolution to produce Lindbladian evolution of the density matrix. In Section~\ref{sec:SI}, we show how our exact RG reproduces the equations for quantum walks. Then, we derive Stochastic Inflation and its NLO corrections from our master equation, and discuss how these techniques can be applied at NNLO and beyond. In Section~\ref{sec:RG}, we demonstrate that dynamical RG in dS resums all UV logs. We conclude in Section~\ref{sec:conclusions}.

\section{Quantum Walks for Cosmology}\label{sec:walks}

The evolution of a scalar field in de Sitter has long been argued to (locally) resemble a random walk~\cite{Vilenkin:1983xq,Starobinsky:1986fx}. The resulting stochastic approach is classical in nature, leaving quantum mechanics responsible only for the microscopic origin of the noise. The universe, of course, remains quantum mechanical and ultimately the classical random walk must be replaced with a quantum generalization. In this section, we will present some simple quantum walks that will serve as useful intuition for de Sitter space. The models in the section are closely related to the literature on quantum Brownian motion~\cite{PhysRevLett.46.211,Grabert:1988yt,Hu:1991di}, however it is important that the quantum walks relevant to cosmology are not equivalent to these models and we will emphasize the points of departure.

\subsection{Density Matrix of a Gaussian Walk}

We will imagine our system is split into a massive field whose wavefunction in the position basis, $\Psi(X)$ encodes the random fluctuations of a $N$ harmonic oscillators in their ground states,
\beq
\psi(\x_i) =\frac{1}{(2\pi\sigma^2)^{1/4}} \exp\left(-x_i^2/(4\sigma^2)\right) \ . 
\eeq
In this regard, the location of heavy field, $X$, defines a pointer basis~\cite{Zurek:1981xq} for the measurement of the oscillators. We can implement a classical gaussian random walk using this quantum system as follows: for $N =0$, the initial state of the walker is definitely at the origin, $\Psi(X) = 
\delta(X)$. We then act on $|\Psi\rangle$ by a shift proportional to $x_i$ for each $i = 1,..,N$:
\beq\label{eq:class_gauss}
|\Psi,N\rangle = e^{i \hat P \sum_i \hat x_i} |X=0\rangle \otimes \prod_i |\psi_i \rangle \ .
\eeq
The reduced density matrix for the walker after coupling to $N$ oscillators, is given by 
\bea\label{eq:X_reduced}
{\rm Tr}_{\psi_i}\rho = \rho_N(X,X') &=&  \left(\prod_i \int dx_i |\psi(x_i)|^2 \right) \left|X =\sum_i x_i \right\rangle \left\langle X' = \sum_i x_i \right|\\
&=& \delta(X-X') \frac{1}{\sqrt{2\pi \sigma^2 N} } \exp( - X^2/(2 \sigma^2 N) ) \ . \label{eq:GQW}
\eea
The second line can be determined by solving Equation~(\ref{eq:class_gauss}) in the momentum basis of the walker. This is mathematically equivalent to a common derivation of probability distribution for $N$-steps of a classical random walk using the Fourier transform. Indeed, this is a classical random walk for all practical purposes.

To gain intuition for our evolution equations, it is useful to derive the master equation for the evolution of the density matrix when increasing the number of harmonic oscillators  from $N$ to $N+1$,
\beq
\rho_{N+1}(X,X') = {\rm Tr}_{x_{N+1}}\left[ e^{i \hat P \hat x_{N+1}}\rho_N \otimes |\psi\rangle \langle \psi | e^{-i \hat P \hat x_{N+1}}  \right] \ .
\eeq
We can use this expression to derive an equation for the time evolution in terms of $T = N dt$ in the continuum limit $N \to \infty$ and $dt \to 0$ while holding $T$ and $\sigma^2 N = \alpha T$ finite. 
Taking the trace in the $x_{N+1}$-position basis and treating $x_{N+1} ={\cal O}(dt^{1/2})$ we can expand both sides to ${\cal O}(dt)$ to find
\beq\label{eq:rho_heat}
\frac{\partial}{\partial T} \rho = - \frac{\alpha}{2} \left( \hat P^2 \rho + \rho\hat P^2 -2  \hat P \rho \hat P \right) \ .
\eeq
In the $X$ position basis, $\rho(X,X',T)$ this becomes the differential equation
\beq
\frac{\partial}{\partial T}\rho(X,X',T) = \frac{\alpha}{2} \left( \frac{\partial^2}{\partial X^2} +\frac{\partial^2}{\partial X'{}^2}  + 2 \frac{\partial^2}{\partial X \partial X'} \right) \rho(X,X',T)  \ . \label{eq:evoldensity}
\eeq
Naively, one might expect the diagonal elements, $X=X'$, to obey the classical heat equation, but this is not the case. In fact, the density matrix is actually singular if we take $X=X'$, which suggests that we need to be more careful to solving for the density matrix, even when we expect it to be diagonal.

The master equation for our walk is easier to solve in the momentum representation of the density matrix, $X \to P$ and $X' \to P'$, so that our evolution is given by
\beq
\frac{\partial}{\partial T}\rho(P,P',T) = - \frac{\alpha}{2} (P-P')^2 \rho(P,P',T) \ .
\eeq
We can directly solve this first order ODE to find
\beq
\rho(P,P',T) = \rho(P,P',0) \exp\left( -\frac{\alpha}{2} (P - P')^2 T  \right) \ .
\eeq
We can find the solution in the position basis by Fourier transform
\beq
\rho(X,X',T) = \int \frac{dP dP'}{(2\pi)^2} e^{i P X} e^{-i P' X'} \rho(P,P',0) \exp\left( -\frac{\alpha}{2} (P - P')^2 T  \right) \ .
\eeq
If we impose that $\rho(P,P',0) = 1$, directly integrating reproduces Equation~(\ref{eq:GQW}). 

We can gain a lot of useful intuition for these walks rewriting our results in terms of Wigner distribution function\footnote{We can also define the Wigner distribution starting from momentum representation of the density matrix, $\rho(P,P',T)$, by taking the inverse Fourier transform of $P-P' \to x$, while the ``momentum" of the Wigner representation becomes the other linear combination $P + P' \to 2p$.},
\beq
W(x,p) =  \int d\Delta x e^{-i p \Delta x} \rho\left(X=x+ \frac{\Delta x}{2}, X'=x-\frac{\Delta x}{2} \right)  \ .
\eeq
By direct calculation, our solution in Equation~(\ref{eq:GQW}) takes a familiar form,
\beq
W(x,p,T) = \exp(- x^2/(2\alpha T)) \ .
\eeq
We can bring this even closer to the classical description by transforming the master equation for the density matrix, Equation~(\ref{eq:evoldensity}), to an evolution equation for Wigner distribution. Specifically, if we change variabels to $X,X' = x \pm \Delta x$ and using the chain rule
\beq
\frac{\partial}{\partial x} =  \frac{\partial}{\partial X}+ \frac{\partial}{\partial X'}  \ .
\eeq
This is the linear combination of derivatives in the evolution of the density matrix, so that
\beq
\frac{\partial}{\partial T} W(x,p,T) =\frac{\alpha}{2} \frac{\partial^2}{\partial x^2} W(x,p,T) \ .
\eeq
In this precise sense, we see that the Wigner distribution obeys the heat equation for a Gaussian quantum walk in the position basis. This is a strong hint that the Wigner distribution provides a more natural starting point for discussing the evolution of these kinds of quantum walks.

None of the conclusions in this section should be surprising, as we have simply translated a purely classical random walk into quantum language. However, in anticipating the possibility of a more fundamentally quantum system, it is useful to look at mathematically equivalent (and still effectively classical walks) in other representations. For example, suppose our walk was defined in momentum space so that one step of the walk shifts
\beq
|\Psi,N\rangle = e^{i \hat X \sum_i \hat x_i} |P=0\rangle \times \prod_i |\psi_i \rangle \ .
\eeq
Following the above logic with $\hat P \to \hat X$, the density matrix obeys the master equation in the position basis
\beq
\frac{\partial}{\partial T}\rho(X,X',T) = - \frac{\alpha}{2} (X-X')^2 \rho(P,P',T) \ .
\eeq
Of course, in terms of the Wigner distribution, using $X-X' = 2 \Delta x$, this is just the heat equation 
\beq
\frac{\partial}{\partial T} W(x,p,T) =\frac{\alpha}{2} \frac{\partial^2}{\partial p^2} W(x,p,T) \  .
\eeq
In terms of Wigner distribution, both types of walks look purely classical. In contrast, recognizing a classical walk in terms of the density matrix directly requires picking the right variables from the outset.

\subsection{Quantum Brownian Motion with a Potential}

Now we would like to introduce additional forces on our quantum system. We start from the same Gaussian Walk for a quantum particle via $N$ harmonic oscillators in a bath, but now we will additionally we immerse the walker, $X$, in a deterministic potential $V(X)$. As a result, we have the combined Hamiltonian of the walker and the harmonic oscillators as 
\begin{equation}
    H = \frac{P^2}{2M} + V(X) + P \sum_i x_i \ .
\end{equation}
We can understand this Hamiltonian using the operator evolution equations, or the classical equations of motion,
\begin{equation}
    \dot{X} = \frac{P}{M}  +\textbf{} x_i\qquad \dot{P} = -V'(X) \ .
\end{equation}
This model is very similar to the Caldeira-Leggett model~\cite{PhysRevLett.46.211} but we notice that the noise directly in the position variable, rather than a random force that appears in the $\dot P$ equation of motion. If $X \to x$ and $P \to p$ were purely classical statistical variables, we expect the classical phase space probability distribution $W_c(x, p, T)$ to follow the Fokker-Planck equation
\begin{equation}\label{eq:Classical_Brownian}
    \frac{d}{dT} W_c(x, p ,T) = \alpha^2 \frac{\partial^2}{\partial x^2} W_c(x, p ,T) - \frac{p}{M} \frac{\partial}{\partial x} W_c(x, p, T) + V'(x) \frac{\partial}{\partial p} W_c(x, p, T) \ .
\end{equation}
Of course, $X$ and $P$ are operators and require a quantum mechanical equivalent of this evolution equation. After tracing over the harmonic oscillators, $x_i$, the evolution of the walker will be described by the evolution of the reduced density matrix, $\rho_R$. We can write this evolution using the Lindbladian formalism, so that 
\begin{equation}
    \dot\rho_R = i[H, \rho] + \sum_\mu \bigg(L_\mu \rho L_\mu^\dagger - \frac{1}{2}\{L_\mu^\dagger L_\mu, \rho_R \}  \bigg) \ , \\
\end{equation}
where $L_\mu$ are the jump operators given by
\begin{equation}
    L_\mu = i P {}_i \bra {\mu} \hat{x}  \ket{0}_i \ .
\end{equation}
The states $|\mu\rangle_i$ and $|0\rangle_i$ are states of the $i$th harmonic oscillator that we are tracing over at time $t$. The jump operator is non-zero only for the first excited state, 
\begin{equation}
    L_{\mu=1} = i P \alpha \qquad \alpha = \bra{1} \hat{x} \ket{0} = \sqrt{\frac{1}{2 m \omega}} = \sigma \ ,
\end{equation}
and the Lindbladian evolution is therefore
\begin{equation}
\begin{aligned}
    \dot \rho_R &= -i [H, \rho] - \alpha^2(p - p')^2 \rho\\
    &= -i\left( \frac{p^2}{2M} - \frac{p'^2}{2M} \right) \rho - i(V(x) - V(x')) \rho_R  - \alpha^2 (p - p')^2 \rho_R \ .
\end{aligned}
\end{equation}
To compare to the classical probability distribution, we transform to the Wigner representation 
\begin{equation}
    \frac{\partial W}{\partial t} = \alpha^2\frac{\partial^2 W}{\partial x^2} - i \left(V \left(x + \frac{i}{2} \partial_p \right) - V \left(x - \frac{i}{2} \partial
    _p \right) \right)W - \frac{p}{M} \frac{\partial W}{\partial x} \ .
\end{equation}
Expanding around the extremum where $\partial_p$ is small, we get that
\begin{equation}
    V \left(x + \frac{i}{2} \partial_p \right) - V \left(x - \frac{i}{2} \partial
    _p \right) = i \frac{\partial}{\partial p}V'(x) - \frac{i}{24} \frac{\partial^3}{\partial p^3} V'''(x) + \hdots
\end{equation}
and thus we have 
\begin{equation}
    \frac{\partial W}{\partial t} = \alpha^2 \frac{\partial^2 W}{\partial x^2} - \frac{p}{M} \frac{\partial W}{\partial x} + V'(x)\frac{\partial W}{\partial p} - \frac{V'''(x)}{24} \frac{\partial^3 W}{\partial p^3} + \hdots
\end{equation}
Comparing this to the classical phase space distribution \eqref{eq:Classical_Brownian}, we see that we get higher order derivatives in $p$ purely due to a quantum effect. These terms are suppressed by powers of $\hbar$.

\subsection{A walk for stochastic inflation} \label{subsec:model_si}

Having introduced quantum walks with a potential, the final ingredient required for building intuition for cosmology is introduction of friction for the walker, $X$. This is straightforward in the Langevin description of a classical walk but is much more involved for a quantum walk. However, we have seen that quantum walks end up looking like classical walks in phase space when using the Wigner distribution. We will therefore consider a classical walk with friction that will reproduce key features of our quantum walk for the dS density matrix.

In the presence of friction, the general Langevin equation for a classical walker takes the form
\beq\label{eq:general_walk_Langevin}
\dot{x} = \frac{p}{m} + \xi_1 \qquad \dot{p} = - \gamma p - V'(x) + \xi_2 \ ,
\eeq
where $\gamma>0$ controls the friction. For most of Newtonian physics applications, one considers Langevin equations with $\xi_1 = 0$. Such a system of equations describes random forces due to thermal baths, and the equilibrium probability distribution is the Maxwell-Boltzmann distribution
\beq\label{eq:Maxwell_Boltz_distr}
W_{\text{eq}}(x, p) \propto \exp \left(-\beta \left( \frac{p^2}{2m} + V(x) \right) \right) \ .
\eeq
This is not the kind of walks that describes fields in de Sitter. Instead, we should consider a  Langevin system where the noise is only in $x$ rather than in $p$, $\xi_1 \neq 0$ and $\xi_2=0$. In other words, we want to study the following walk
\beq
\dot{x} = \frac{p}{m} + \sqrt{2D}\xi \qquad \dot{p} = -\gamma p - V'(x) \ ,
\eeq
which represents a particle in an external potential, with friction in $p$ characterized by $\gamma$ and kicks in $x$ characterized by $D$. As we will see, even at $\hbar \to 0$, this equation reproduces key features that we will encounter in stochastic inflation. The phase space probability distribution, $W_c$, at leading order, evolves as
\begin{equation}\label{eq:FPl_for_stoch_walk}
    \frac{dW_c}{dt} = D\frac{d^2W_c}{dx^2} + \frac{d}{dp} (\gamma p W_c + V'(x)W_c) - \frac{p}{m} \frac{dW_c}{dx} \ .
\end{equation}
In order to solve this system, we will consider the overdamped limit $\gamma \gg 1/t$, and will derive the solutions as a series in this limit. We assume that at leading order, the background value $x$ is fixed while $p$ fluctuates to come to equilibrium around a given value of $x$. The expectation value of $p$ is described by
\begin{equation}
    \begin{aligned}
        \frac{d}{dt} \vev{p}_x &= -\gamma \vev{p}_x - V'(x) \ .
    \end{aligned}
\end{equation}
The solution is
\begin{equation}
    \vev{p}_x = e^{-\gamma t} \vev{p(0)}_x - \int_0^t ds\, e^{\gamma (s- t)} V'(x) = e^{-\gamma t}\vev{p(0)}_x - \frac{V'(x)}{\gamma} ( 1 - e^{-\gamma t} ) \ ,
\end{equation}
where we are treating $V(x)$ as time independent at leading order. We see that in the overdamped limit, $\vev{p}_x \to -V'(x)/\gamma$. Now, we can solve for the probability distribution of $x$. Integrating over $p$ and substituting $\langle p \rangle_x$, we get the Fokker-Planck equation for $x$
\begin{equation}\label{eq:Fokker_pl_cw}
    \frac{d P(x)}{dt} = D\frac{d^2 P(x)}{d x^2}  + \frac{d}{dx} \left(\frac{V'(x)}{m \gamma} P(x) \right) \ ,
\end{equation}
and hence the equilibrium distribution for $x$ given by
\begin{equation}
    P_{eq}(x) = A e^{-V(x)/m \gamma D} \ .
\end{equation}

We would also like to construct the equilibrium solution for the full phase space distribution, i.e., the equilibrium solution for \eqref{eq:FPl_for_stoch_walk}. However, unlike the case for the Gibbs distribution, we do not have an analytic solution for generic $V(x)$. However, a closed form solution does exist for quadratic $V(x)$. Writing the potential as $V(x) = \frac{1}{2} k x^2$, we see that the equilibrium probability distribution is given by
\begin{equation}
    W_{eq} = \alpha \exp(Ax^2 + Bxp + C p^2) \ ,
\end{equation}
where 
\begin{equation}
    A = -\frac{\gamma}{2D} \qquad B = -\frac{\gamma^2}{kD} \qquad C = -\frac{\gamma}{2mkD} - \frac{\gamma^3}{2k^2D} \ .
\end{equation}
Explicitly, we have 
\begin{equation}\label{eq:eq_dist_x_p}
    W_{eq} \propto \exp \left( -\frac{\gamma}{2D} x^2 - \frac{\gamma^2}{D k}xp - \frac{\gamma}{2 kD} \left(\frac{1}{m} + \frac{\gamma^2}{k} \right) p^2 \right) \ .
\end{equation}
On integrating out $p$, we get
\begin{equation}
    P(x) = A\exp \left( - \frac{\frac{1}{2}k\gamma}{k D + D \gamma^2 m} x^2 \right) \ .
\end{equation}
Writing this in the form $P \propto \exp(-V_{\rm eff}(x)/m \gamma D)$, we see that the effective potential is given by 
\begin{equation}\label{eq:veff_harmonic}
    V_{\rm eff} = \frac{1}{2}kx^2 - \frac{1}{2}\frac{k^2  x^2}{m\gamma^2} + \cdots
\end{equation}
where the expansion is in $k/m \gamma^2 \ll1$. We see that even in the case of a Gaussian theory, the momentum will induce a non-trivial effect on the non-perturbative probability distribution of the position, and in particular on the tails. Finally, we note that our equilibrium solution is quite different from the thermal Gibbs distribution, which cleanly factorizes out as $\exp(-\beta(p^2/2m + V(x)))$. We will comment more on this in the next sections.

\subsubsection*{Corrections to the effective potential} 

Having provided the leading order solution for the evolution equation for $P(x)$, we want to understand the next order correction. We can do this in the Langevin picture
\begin{equation}
    \dot{p} = -\gamma p - V'(x) \qquad \dot{x} = \frac{p}{m} + \sqrt{2D}\xi(t) \ .
\end{equation}
Formally, we can solve this via
\begin{equation}
\begin{aligned}
    p(t) &= -\int ds e^{-\gamma(t-s)} V'(x(s))\\
    &= -\int ds e^{-\gamma(t - s)} V'(x(t)) - \int ds e^{-\gamma(t - s)}(s-t) V''(x(t)) \dot{x}(t) + \cdots
\end{aligned}
\end{equation}
Using the Langevin equations to substitute $\dot{x}$ in terms of $p$, we get
\begin{equation}
p = -\frac{V'(x)}{\gamma} - \frac{V'(x) V''(x)}{m\gamma^3} + \frac{V''(x)}{\gamma^2} \sqrt{2D} \xi(t)  + \cdots
\end{equation}
So, substituting this back into the equations for $x$, we get
\begin{equation}
    \dot{x} = \sqrt{2D} \xi(t)\left(1 + \frac{V''(x)}{m\gamma^2} \right) - \left(\frac{V'(x)}{m\gamma} + \frac{V'(x) V''(x)}{m^2\gamma^3} \right) \ .
\end{equation}
Using the Stratonovich convention, we see that this translates into the Fokker-Planck equation
\begin{equation}
    \frac{\partial P}{\partial t} = D \frac{\partial}{\partial x} \left( \left(1 + \frac{V''}{m\gamma^2} \right)\frac{\partial}{\partial x} \left(1 + \frac{V''}{m\gamma^2} \right) P(x)\right) + \frac{\partial}{\partial x} \left( \frac{V'(x)}{\gamma} + \frac{V'(x) V''(x)}{m\gamma^3} \right)P \ .
\end{equation}
For the harmonic case, we get that
\begin{equation}
    \frac{\partial P}{\partial t} = D\left(1 + \frac{k}{m\gamma^2}\right)^2 \frac{\partial^2P}{\partial x^2} + \frac{\partial}{\partial x} \left[\left( \frac{k x}{m\gamma} + \frac{k^2 x}{m^2 \gamma^3} \right)P\right] \ ,
\end{equation}
For a Fokker-Planck equation of the form \eqref{eq:Fokker_pl_cw}, the equilibrium solution is given by
\begin{equation}
    P_{\rm eq} \propto \exp\left(-\frac{V_{\rm eff}}{m \gamma D} \right) \ .
\end{equation}
Thus, for the harmonic case, since we have that our effective diffusion in the Fokker-Planck equation is $D' = D\left(1 + \frac{k}{m\gamma^2}\right)^2 $, we get that $V_{\rm eff}$ is given by
\begin{equation}
    V_{\rm eff} = \frac{1}{2} kx^2 \left(1 + \frac{k}{m \gamma^2} \right)\left(1 - 2\frac{k}{m \gamma^2} \right) = \frac{1}{2} k x^2 \left(1 - \frac{k}{m \gamma^2} \right) \ ,
\end{equation}
which matches the harmonic effective potential in \eqref{eq:veff_harmonic}. In section \ref{sec:SI}, we will use a morally equivalent procedure to derive similar corrections to stochastic inflation.

\subsubsection*{Comments on the equilibrium distribution}

While we have a stationary (time independent) solution, it is crucially different from the Gibbs distribution, which exhibits detailed balance. Detailed balance tells us that there are no probability currents, and that forward and backward transition rates are equally likely 
\begin{equation}
    \frac{dP_{i}}{dt} = \sum W(j \to i)P_j - \sum W(i \to j)P_i = 0 \qquad W(j \to i) P_j = W(i \to j) P_i \,\,\forall i,j
\end{equation}
This is the statement of microscopic time reversibility. However, for our stationary solution, time reversibility and hence detailed balance is manifestly broken, and we have non-zero currents which tell us the preferential probability flow. Writing
\begin{equation}
    \frac{dP}{dt} =- \nabla \cdot J = -(\partial_x J_x + \partial_p J_p)  \ ,
\end{equation}
we get 
\begin{equation}
    J_x = \frac{p}{m}W - D \frac{\partial W}{\partial x} \qquad J_p = -\gamma p W - V'(x) W \ .
\end{equation}
Evaluating the currents for our equilibrium solution in equation \eqref{eq:eq_dist_x_p}, we get 
\begin{equation}
    J_x = \left( \gamma x + \frac{\gamma^2p}{k} + \frac{p}{m}\right) W_{eq} \qquad J_p = - \gamma \left( p + \frac{V'(x)}{\gamma} \right) W_{eq} \ .
\end{equation}
We can rewrite this, in our small parameter limit $k \ll m \gamma^2$, as 
\begin{equation}
    J_x = \frac{\gamma^2}{k} \left( p + \frac{V'(x)}{\gamma} \right) W_{eq} \qquad J_p = - \gamma \left( p + \frac{V'(x)}{\gamma} \right) W_{eq} \ .
\end{equation}
We thus see that whenever we are off the momentum saddle point axis $p = -V'(x)/\gamma$, we get a non-zero current. However, on integrating out $p$, we constrain ourselves to flow along this axis. Thus, the reduced one dimensional Fokker-Planck equation
\begin{equation}
    \frac{dP}{dt} = D \frac{d^2P(x)}{dx^2} + \frac{d}{dx} \left(\frac {V'(x)}{m \gamma} P(x) \right) 
\end{equation}
does satisfy detailed balance at equilibrium, as expected. Thus, stochastic inflation represented in the flat FRW slicing has an arrow of time given by non-zero currents when the observer has access to the full phase space distribution. 

\section{Exact RG and the Density Matrix}\label{sec:exact}

We would now like to take our intuition from quantum walks and apply it to the reduced density matrix of the long wavelength modes in de Sitter. Specifically, we will be constructing a density matrix
\beq
\rho_R[\varphi_+(\x) ,\varphi_-(\x),t] = {\rm Tr}_{\Phi} \rho[\phi_+,\phi_-]  \qquad \phi_\pm= \varphi_\pm+ \Phi
\eeq
where $\varphi(\x,t)$ is a local field defined in terms of the Fourier modes $\phi(\k,t)$ with $k < aH$ and $\Phi(\x,t)$ is the local field composed of the hard-modes with $k\geq aH$ that we want to trace over.

We have intuition from RG flow and EFT that suggests that the evolution of the reduced density matrix can be understood as the exact RG for the density matrix $\rho[\phi_+,\phi_-]$ after integrating our a shell of modes with $k \in [\Lambda, \Lambda+\delta \Lambda]$ where $\Lambda > aH$. For purely long wavelength observables, we are going to trace over all the short modes anyway and $\rho_R$ is independent of $\Lambda \gg aH$, and plays the same role as the Euclidean partition function of an exact RG in flat space~\cite{Polchinski:1983gv}. However, because of the expansion of the Universe, modes that are in $\rho_R(t)$ would not have been in $\rho_R(t')$ for $t' \ll t$. Therefore, understanding dynamical RG (RG plus time evolution) introduces subtleties beyond the usual exact RG. Nevertheless, we will find that the by following analogous exact RG treatment for the density matrix in flat space~\cite{Goldman:2024cvx,Green:2024cmx} (see also~\cite{Balasubramanian:2014bfa,Cotler:2022fze}), we can derive the evolution of the density matrix in dS under time evolution. Our approach is related to previous work deriving the Stochastic formalism using functional/exact RG~\cite{Gautier:2013aoa,Guilleux:2015pma,Prokopec:2017vxx,Cespedes:2023aal}.

\subsection{Setup}

We will start from the density matrix associated with the UV field $\phi$, $\rho[\phi_+,\phi_-]$. This field is defined up to some physical cutoff $\Lambda_0$. Our interest is not the UV density matrix, but instead the reduced density matrix of the long distance modes. We can do this at the level of the path intergal by integrating out shells of momentum down to an effective cutoff $aH < \Lambda < \Lambda_0$. The density matrix of the IR field, $\vp$, can be defined a path integral over the contour $\gamma \in (-\infty(1-i\epsilon), t_0] \cup [t_0,-\infty(1+i\epsilon)$ with the boundary condition $\vp(\x,t_\pm\to t_0) = \vp_\pm(\x)$ where $t_+ \in (-\infty(1-i\epsilon), t_0]$ and $t_- \in  [t_0,-\infty(1+i\epsilon)]$. We then express the density matrix in terms of the action as
\beq\label{eq:action}
\begin{aligned}
\rho_R[\vp_+,\vp_-] & =\int_\gamma D \varphi T \exp \left(i S_0+i S_{\rm int}  + i S_\partial \right)\\
S_0[\vp(\vec{x}, t), \Lambda] & \equiv \int_\gamma d t a^3(t) \int \frac{d^3 k}{(2 \pi)^3} \frac{1}{2} \nabla_\mu \vp(\k,t) K^{-1}\left(\frac{k^2}{\Lambda^2}\right) \nabla^\mu \vp(\x,t) 
\end{aligned}
\eeq
where $\Lambda$ is the scale where we integrate out the short modes to define an EFT. The UV density matrix, $\rho[\phi_+, \phi_-]$ is defined in the same way, but with $\Lambda=\Lambda_0$. Our goal is to show that in integrating out a shell $[\Lambda, \Lambda + \Delta \Lambda]$, the form of the density matrix does not change.

We will define $\nabla_\mu$ is the covariant derivative and $\nabla^2 = \nabla_\mu \nabla^\mu$. To simplify notation, when acting on Fourier modes, it is implied that we also Fourier transformed the derivative, so that (for example)
\beq
\nabla_\mu \nabla^\mu f(\k,t) \equiv \left(\frac{\partial^2}{\partial t^2} + 3 H \frac{\partial}{\partial t} + \frac{k^2}{a^2} \right) f(\k,t) \ .
\eeq
The additional terms in the action $S_I$ and $S_\partial$ and the bulk interaction and boundary terms respectively. The reduced density matrix will then be traced over all modes with $k > aH$ so we will have $\Lambda \geq aH(t)$.

For the density matrix, a boundary term is always required in order to have a well-defined action principle. At the quadratic level, this implies 
\beq
S_{0,\partial} \supset -\int d^3 x a^3(t) \frac{1}{2} \left( \vp_+ K^{-1} \dot \vp(t_{0,+}) - \vp_-  K^{-1} \dot \vp(t_{0,-})\right) \ ,
\eeq
where $t_{0,\pm}$ means we evaluate the derivative with respect to $t_\pm$ and then take $t_\pm \to t_0$.

\subsection{Integrating Out}

Now we will proceed to integrate out a shell of momentum. Because we are performing the path integral over $t< t'$, any change to the density matrix as we change $\Lambda$ must arise from fields at time $t$. We will show that all such effects can be captured by the boundary action, $S_\partial$ so that
\beq\label{eq:exactRG_def}
\frac{\partial}{\partial s } \rho_R[\varphi_+ ,\varphi_-,t]=i\int_\gamma D \varphi   \left(\frac{\partial}{\partial s }S_0+\frac{\partial}{\partial s } S_I  + \frac{\partial}{\partial s } S_\partial \right)\exp \left(i S_0+i S_{\rm int}  + i S_\partial \right) = 0 \, ,
\eeq
where $s \equiv \log \Lambda$. As in the case of exact RG, we are looking to define the change to the interaction and boundary terms that will make the vanishing of this expression manifest. At this stage, Equation~(\ref{eq:exactRG_def}) is a conjecture, as we must demonstrate that any dependence of the density matrix on $\Lambda$ can be absorbed into $S_\partial$.

The key idea is to use our knowledge of the quadratic (bulk) action to simplify this expression. In order to simplify the notation, we will suppress the integral over Fourier modes and isolate the integration over time. Quantities that are integrated over same time variable we will write in the compact notation
\beq
f*g \equiv \int dt f(t)g(t) \qquad f*H*g \equiv \int dt \int dt' f(t) H(t,t') g(t') \ .
\eeq
We will keep track of time integration in this way in order to isolate the effect of the non-trivial boundary conditions imposed at $t = t_{0,\pm}$. 

\subsubsection*{Quadratic Terms}

Now we can use our knowledge of the free theory to write
\beq\label{eq:S0def}
\frac{\partial}{\partial s }S_0 = \frac{1}{2} \frac{\delta}{\delta \varphi} S_0 * G' * \frac{\delta}{\delta \varphi} S_0  + \delta S_{\partial, 2} \ , 
\eeq
where we defined the Green's function $G(t,t')$ so that
\beq
G^{-1} = K^{-1} \nabla_\mu \nabla^\mu \qquad G^{-1} G  =a^{-3}(t) \delta(t-t')\delta(\x-\x')  \qquad G' =\frac{\partial}{\partial s } G = \left(\frac{\partial}{\partial s } K \right)  G \ .
\eeq
The boundary term $\delta S_{\partial, 2}$ is determined using integration by parts. Specifically, we can write
\beq\label{eq:pS0}
\frac{\partial}{\partial s }S_0 =   a^3(t) \frac{K'}{K^{2}}  \vp \nabla^2 \vp = (K^{-1} \vp \nabla^2) * G' *( K^{-1} \nabla^2 \vp) \ ,
\eeq
using the definition of the Green's function. The correction to the quadratic boundary term comes from integrating Equation~(\ref{eq:pS0}) by parts until it takes the form of Equation~(\ref{eq:S0def}). Defining $\bar S_0 =S_0 + S_{0,\partial}$ and using 
\beq
\frac{\delta}{\delta \vp} \bar S_0=K^{-1} a^3(t) \nabla^2  \vp \ ,
\eeq
we can see that 
\beq
 \frac{\delta}{\delta \vp} \bar S_0 * G' * \frac{\delta}{\delta \vp} \bar S_0  = K^{-1} \vp \nabla^2 * G' * K^{-1} \nabla^2 \vp - K^{-1} (\nabla_0 \vp G' -\vp\nabla_0 G')|_{t_{0,-}}^{t_{0,+}} * \frac{\delta}{\delta \vp} \bar S_0 \ .
\eeq
As written yet, this is not a boundary term because it still involves a time integral through the $\delta_\vp \bar S_0$ term. Fortunately we can integrate the second term  by parts, and after simplification we find that only a boundary term remains of the form  $\delta S_{\partial,0} = \int d^3 x a^6(t_0) K^{-2} \frac{1}{2} f(\vp)$ where 
\begin{align}
f & =(\vp_- -\vp_+)\left(\nabla_t G^{\prime}\left(t_{+}\right) \nabla_{t^{\prime}} \vp\left(t_{-}\right)- \nabla_t G^{\prime}\left(t_{-}\right) \nabla_{t^{\prime}} \vp\left(t_{+}\right) 
 -\nabla_t \nabla_{t^{\prime}} G^{\prime} (\vp_- -  \vp_+)\right) \\
 & -(\nabla_t \vp(t_-) -\nabla_t \vp(t_+)) \left( G^{\prime} ( \nabla_t \vp(t_-)- \nabla_t \vp(t_+))  -\nabla_{t^{\prime}} G^{\prime}\left(t_+\right) \vp_- + \nabla_{t^{\prime}} G^{\prime}\left(t_{-}\right) \vp_+ \right) \nonumber
\end{align} 
In writing this expression, we anticipated that 
\beq
i G'(t,t',\p)|_{t=t'=t_0}  \equiv \Delta_0(p) \qquad i \nabla_t \nabla_{t'} G(t,t',\p)_{t=t'=t_0} \equiv \Delta_2(p)
\eeq
are independent of take $t_\pm \to t_0$. To determine the remaining terms, we note that $G(t,t',\p)$ is the path ordered Green's function along $\gamma$,
\beq
G'(t,t',\p) = i \langle {\rm T}_\gamma \vp(\p,t_1) \vp(-\p,t_2)\rangle' = i \frac{dK}{ds} \frac{H^2}{2p^3} (1-i p \tau_1)(1+i p \tau_2)e^{i p (\tau_1 -\tau_2)}
\eeq
where $t_1$ ($t_2$) is the early (later) time along the contour $\gamma$ and $\tau_i$ is the conformal times associated with physical time $t_i$. As a result, we have, for the massless field
\bea
 \Delta_0(p,t_0) &=&iG' = iG'(t_0,t_0,\p) =  -\frac{\partial K}{\partial s} \frac{H^2}{2p^3} (1+ \tau_0^2 p^2)\\
\Delta_2(p,t_0)&=& \partial_t\partial_{t'} i G'(t_0,t_0,\p) = -\frac{\partial K}{\partial s} \frac{H^4}{2p^3} \tau_0^4 p^4 \ .
\eea
The single derivatives depend on the order, and the order of limits that we take as $t, t' \to t_0$ matters, and we find 
\bea
i\partial_t G'(t,t_0,\p)|_{t=t_\pm} = i\partial_t' G'(t_0,t',\p)|_{t'=t_\mp} = - \frac{\partial K}{\partial s} \frac{H^3}{2p^3} (\tau_0^2 p^2)(1\pm i \tau_0 p ) \equiv  \Delta_R \pm i \Delta_I \ .
\eea
where 
\bea
\Delta_{R} &= - K' \frac{H^{3}}{2p^{3}}(p^{2} \tau_{0}^{2}) \qquad i\Delta_{I} = i K^{-1} K' \frac{H^{3}}{2p^{3}}(\tau_{0}^{3} p^{3}) = -i \frac{K'}{2a^{3}}
\eea
The final result is that 
\begin{align}
    i f =& -\Delta_2 (\vp_- -\vp_+)^2   - \Delta_0 (\nabla_t \vp(t_-) -\nabla_t \vp(t_+))^2\\
    &+ \Delta_R \{(\vp_+-\vp_-),(\nabla_t \vp(t_-) -\nabla_t \vp(t_+)) \} \nonumber\\
    &+i\Delta_{I} \bigg[(\vp_{-} - \vp_{+})(\nabla_{t} \vp_{-} + \nabla_{t} \vp_{+}) - (\nabla_{t} \vp_{-} - \nabla_{t} \vp_{+} )(\vp_{+} + \vp_{-})\bigg] \nonumber \ .
\end{align}
These additional quadratic terms will play a significant role in stochastic inflation. 

The corrections to the boundary action will be given two different interpretations but are equivalent. The first, in precise analogy with exact RG, is that we change the quadratic boundary action to absorb this term
\beq
\frac{\partial}{\partial s} S_\partial \supset - i f \ .
\eeq
This interpretation is most useful when discussing the resummation of logs in the sense of RG flow.  Although this is a local boundary term, it is unfortunately not equivalent to unitary evolution. Specifically, we have mixed $\vp_+ - \vp_-$ terms that we associate with open quantum systems. The second, and more typical, interpretation of the boundary terms is to treat them terms as a change to the density matrix,
\beq
\frac{\partial}{\partial s} \rho_{R} \supset if \rho_{R} \ .
\eeq
Here $if$ is a local super-operator in the $\vp_\pm$ basis. This interpretation will be most useful for describing the density matrix during stochastic inflation. 

Finally, the running of the required quadratic boundary terms of the action will also contribute a correction to the density matrix via,
\begin{equation}
   \frac{\partial}{\partial s} S_{0, \partial} = a^3(t) \frac{K^{-2}}{2} \left(\frac{\partial K}{\partial s} \right) (\vp_+ \nabla_t \vp_+ - \vp_- \nabla_t  \vp_- ) \ .
\end{equation}
Using the observation that 
\begin{equation}\label{eq:total deriv}
    \Delta_I = \frac{H^3 \tau_0^3}{2} \frac{\partial K}{\partial s} = -\frac{\partial K}{\partial s} \frac{1}{2a^3}  \ ,
\end{equation}
we can rewrite
\begin{equation}
    i\frac{\partial}{\partial s} S_{0, \partial} = a^6(t) \Delta_I K^{-2}(\vp_+ \nabla_t \vp_+ - \vp_- \nabla_t \vp_- ) \ .
\end{equation}
Since $[\vp, \nabla_t \vp] \neq 0$ we need to prescribe a time ordering. The correct time ordering is the symmetrized one, which gives 
\begin{equation}
    i\frac{\partial}{\partial s} S_{0, \partial} = a^6(t) \Delta_I \frac{K^{-2}}{2} (\vp_+ \nabla_t \vp_+ + \nabla_t \vp_+ \vp_+ - \vp_- \nabla_t \vp_- -  \nabla_t \vp_- \vp_- ) \ .
\end{equation}
Now we can combine all the terms to find  
\begin{equation}
\begin{aligned}
    \bigg[\frac{1}{2}if + i K^2\frac{\partial}{\partial s} S_{0, \partial} \bigg] &= -\frac{1}{2}\Delta_2 (\vp_- -\vp_+)^2 - \frac{1}{2}\Delta_0 (\nabla_t \vp(t_-) -\nabla_t \vp(t_+))^2 \\
    &+ \frac{1}{2}\Delta_R \{(\vp_+-\vp_-),(\nabla_t \vp(t_-) -\nabla_t \vp(t_+)) \} \nonumber\\
    &- i\Delta_I (\nabla_t \vp(t_-) - \nabla_t\vp(t_+))(\vp_+ + \vp_-) \ .
\end{aligned}
\end{equation}
Overall, we find that the evolution of the density matrix due to the quadratic terms takes the form
\begin{equation}
    \frac{\partial}{\partial s} \rho_{R} \supset ia^6 K^{-2}\frac{f}{2} \rho_{R} + \frac{\partial S_{0, \partial}}{\partial s} \ ,
\end{equation}
where we have suppressed the implicit momentum integral.

\subsubsection*{Interaction Terms}

After determining the quadratic boundary terms, we then proceed to use the Schwinger-Dyson (SD) equations to eliminate $S_0$. Specifically, at the operator level, the SD equations reflect our ability to integrate by parts so that 
\beq
i \int_\gamma D \varphi   \left ({\cal O} \frac{\delta}{\delta \vp }S_0  \right) e^{i \left(S_0+S_I  \right)}  =-  \int_\gamma D \varphi \left(i{\cal O} \frac{\delta}{\delta \vp} S_I + \frac{\delta}{\delta \vp}{\cal O} \right) e^{i \left(S_0+S_I  \right)} \ ,
\eeq
where $S_I = S_{\rm int} + S_\partial$. This equation holds for $\varphi(t\neq t_0)$ because we are integrating over all fields up to the boundary conditions at $t_0$. One should worry about the potential contributions at $t=t_0$; fortunately these are manifestly pure boundary terms and can be determined directly.

At this stage, there is no practical difference between the dS and the flat space in-in or density matrix calculations~\cite{Goldman:2024cvx,Green:2024cmx}, other than the precise form of $G$. Therefore, we apply the SD equations repeatedly, to write
\beq
\begin{aligned}
\frac{1}{2} \frac{\delta S_0}{\delta \varphi} \star G' \star \frac{\delta S_0}{\delta \varphi}  =&-\frac{1}{2} \int_\gamma d t\left(\frac{\delta^2 S_0}{\delta \varphi \delta \varphi} G'(t, t)\right)+\frac{1}{2} \frac{\delta S_0}{\delta \varphi} \star G' \star \frac{\delta S_I}{\delta \varphi} \\
 =&\frac{1}{2} \int_\gamma d t\left(\frac{\delta^2 S_I}{\delta \varphi \delta \varphi} G'(t, t)\right)-\frac{1}{2} \frac{\delta S_I}{\delta \varphi} \star G' \star \frac{\delta S_I}{\delta \varphi}\\
&-\frac{1}{2} \int_\gamma d t\left(\frac{\delta^2 S_0}{\delta \varphi \delta \varphi} G'(t, t)\right) \ .
\end{aligned} 
\eeq
The term on the last line is a constant and just contributes to the overall normalization of $\rho$ (which is fixed, in principle, by the trace condition, ${\rm Tr}\rho =1$).  What remains are the two (well-known) contributions to the effective action from integrating out a shell of momentum, namely
\beq
\frac{\partial}{\partial s} S_{I} = - \frac{1}{2} \int_\gamma d t\left(\frac{\delta^2 S_I}{\delta \varphi \delta \varphi} G'(t, t)\right)+\frac{1}{2} \frac{\delta S_I}{\delta \varphi} \star G' \star \frac{\delta S_I}{\delta \varphi}
\eeq
The first term is purely a local function of $t$ because there is a single integral and therefore contributes only to $S_{\rm int}$. The second term includes both local and boundary terms. Specifically, because $G'(t,t)$ is exponentially suppressed in $t-t'$ (or, as we will see, $t-t_0$ or $t'-t_0$), we can Taylor expand and integrate to generate local operators. 

\begin{figure}[h!]
	\centering
    \includegraphics[width=0.65\textwidth]{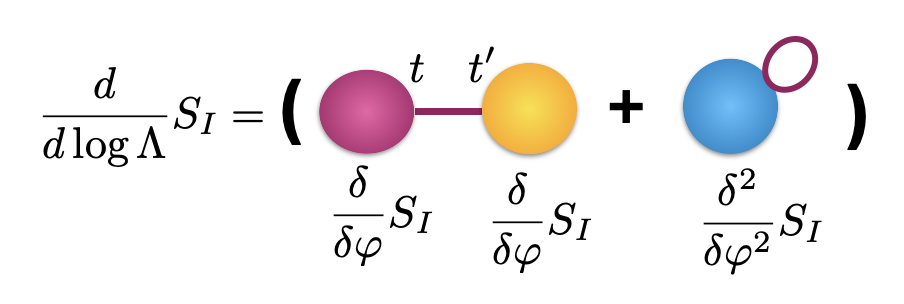}
	\caption{The equation for the evolution of the effective action under dynamical RG. The key difference between this diagrammatic representation and conventional exact RG is that the times $t$ and $t'$ are no necessarily on the same branch of the Schwinger-Keldysh contour, giving rise to additional boundary terms.}
\label{fig:exactRG}
\end{figure}

However, since the local action can only involve operators in one of the $\gamma_+ = t_+\in (-\infty(1-i\epsilon),t_0]$ or $\gamma_-= t_-\in [t_0,-\infty(1+i\epsilon)]$ regions, we may still have terms mixing $t_+$ and $t_-$ that do not arise in the bulk action, see Figure~\ref{fig:exactRG} for illustration. Therefore, defining
\beq
S_{\rm int, \pm} = \int_{\gamma_+} dt_\pm {\cal L} \ ,
\eeq
the correction to the bulk (local) action is
\beq
\frac{\partial}{\partial s} S_{\rm int,\pm}=   - \frac{1}{2} \int_\gamma d t\left(\frac{\delta^2 S_{\rm int,\pm}}{\delta \varphi \delta \varphi} G'(t_\pm, t_\pm)\right)+\frac{1}{2} \frac{\delta S_{\rm int,\pm}}{\delta \varphi} \star G_{\pm,\pm}' \star \frac{\delta S_{\rm int,\pm}}{\delta \varphi} \ . 
\eeq
Allowing for a boundary action $S_\partial(\vp_+,\vp_-)$, we then find boundary terms 
\beq\label{eq:boundary_Sint}
\frac{\partial}{\partial s} S_{\partial,{\rm int}} =  \frac{\delta S_{\rm int,+}}{\delta \varphi} \star G_{+,-}' \star \frac{\delta S_{\rm int,-}}{\delta \varphi} 
\eeq
It will generally be the case that $G_{+,-}$ is exponentially suppressed in $t_{\pm} -t_0$ and therefore
\beq
\frac{\partial}{\partial s} S_{\partial,{\rm int}} = \int d^3 x \sum_{i,j} {\cal O}_{+,i} {\cal O}_{-,j}  \ ,
\eeq
where ${\cal O}_\pm$ are local operators of $\vp(t_\pm)$ defined at $t_\pm  \to t_0$ respectively that can be derived from the Taylor expansion of Equation~(\ref{eq:boundary_Sint}) at $t_+=t_- =t_0$. We see this term necessarily mixes $\vp(t_\pm)$ and therefore is not associated with conventional Hamiltonian evolution. However, because this is a pure boundary term, it can be interpreted as a redefinition of the operators at $t_0$. 

Like the quadratic boundary terms another way to interpret mixed boundary terms is through the evolution of the density matrix. Specifically, $S_\partial$ acts as a super-operator on density matrix and therefore can be incorporated through the master equation
\beq\label{eq:super_int}
\frac{\partial}{\partial s} \rho_R \supset \sum_{i,j} \int d^3 x \, {\cal O}_{+,i} \rho_R {\cal O}_{-,j} \equiv \$_{\rm int} \rho_R \ ,
\eeq
where $\$_{\rm int}$ is the superoperator in the Kraus representation. In this representation, all corrections to the effective action in $S_{\rm int}$ can be derived from any observable as they do not depend on $t_0$ or the fact this is an in-in correlator. All terms specific to the in-in correlator at $t_0$ contribute to the evolution of the density matrix. 

While the presence of the additional boundary terms will be important for cosmological applications, such as stochastic inflation, it is important to note that they are needed in flat space when matching correlators~\cite{Green:2024cmx}. The fact that they do not appear in matching S-matrix elements is a consequence of the fact that they can represent field redefinitions between the UV and IR descriptions.

\subsubsection*{Full Density Matrix RG}

To summarize the exact RG calculation, integrating out a shell of momenta introduces a differential change to the reduced density matrix of the long wavelength modes,
\begin{align}
    \frac{\partial \rho_R}{\partial s} &= \int \frac{d^3p}{(2\pi)^3}\, a^6(t) i K^{-2} \bigg[\frac{1}{2} f(\vp) \rho_R + \frac{\partial S_{0, \partial}}{\partial s}\bigg] \nonumber \\ 
    &= \int \frac{d^3p}{(2\pi)^3}\, a^6(t) K^{-2}  \, \bigg[-\frac{1}{2}\Delta_2 (\vp_- -\vp_+)^2 - \frac{1}{2}\Delta_0 (\nabla_t \vp(t_-) -\nabla_t \vp(t_+))^2 \\
    &+ \frac{1}{2}\Delta_R \{(\vp_+-\vp_-),(\nabla_t \vp(t_-) -\nabla_t \vp(t_+)) \} - i\Delta_I (\nabla_t \vp(t_-) - \nabla_t\vp(t_+))(\vp_+ + \vp_-)  \bigg]\rho_R  + \$_{\rm int} \rho_R \  , \nonumber
\end{align}
where $\$_{\rm int}$ are higher order local terms defined in Equation~(\ref{eq:super_int}). We can further simplify the quadratic terms in this equation by recognizing that 
\begin{equation}
    a^3(t) K^{-1}  \nabla_t \vp(t_+) \rho_R =  -i \frac{\delta}{\delta \vp(t_+)} \rho_R \qquad a^3(t) K^{-1} \nabla_t \vp(t_-) \rho_R =   i \frac{\delta}{\delta \vp(t_-)} \rho_R \ , 
\end{equation}
leading to
\begin{align}\label{eq:RG_rhoR_final}
        \frac{\partial \rho_R}{\partial s} &= \frac{1}{2}\int \frac{d^3p}{(2\pi)^3}\, \,  \bigg[\Delta_0 \left( \frac{\delta}{\delta \vp_+} + \frac{\delta}{\delta \vp_-} \right)^2 - a^6 K^{-2}  \Delta_2 (\vp_- -\vp_+)^2  \\
        & + i \frac{a^3}{K}  \Delta_R \bigg\{(\vp_+-\vp_-), \left(\frac{\delta}{\delta \vp_+} + \frac{\delta}{\delta \vp_-} \right) \bigg\} + 2 \frac{a^3}{K}\Delta_{I} \left(  \frac{\delta}{\delta \vp_{+}}  + \frac{\delta}{\delta \vp_{-}} \right)(\vp_{+} + \vp_{-})  \bigg] \rho_R+ \$_{\rm int} \rho_R  \nonumber \ .
\end{align}
Here it will be important that this is a functional equation as we have suppressed the momentum, $\varphi = \varphi(\p)$.

At this point, it is instructive to compare this to the random walks that we have built intuition for in Section \ref{sec:walks}. Schematically, the quadratic terms (i.e.~dropping $\$_{\rm int}$) in our evolution equation are of the form
\begin{equation}
    \begin{aligned}
        \frac{\partial \rho_R}{\partial s} &= \Delta_0 \left( \frac{\partial}{\partial x} + \frac{\partial}{\partial x'} \right)^2 \rho_R + \Delta_2 (x - x')^2 \rho - i \Delta_R \left( \frac{\partial}{\partial x} + \frac{\partial}{\partial x'} \right) (x - x') \rho_R \\
        &  + \frac{\Delta_I}{2} \left( \frac{\partial}{\partial x} + \frac{\partial}{\partial x'} \right)(x + x') \rho_R  \ . \label{eq:final_RG_quad}
    \end{aligned}
\end{equation}
As we commented on earlier, it is easier to gain intuition for random walks in the Wigner phase space formalism. Transforming the above equation to the Wigner representation, we get
\begin{equation}
    \begin{aligned}
        \frac{\partial W}{\partial t} = \Delta_0\frac{\partial^2}{\partial x^2} W + \Delta_2 \frac{\partial^2}{\partial p^2} W + \Delta_R \frac{\partial^2}{\partial x \partial p} W + \Delta_I \frac{\partial}{\partial x} (x W) \ . \label{eq:compare_RW}
    \end{aligned}
\end{equation}
Looking at this equation, we can quickly identify what the different $\Delta$ represent. The $\Delta_0, \Delta_2$ terms represent the diffusion in $x, p$, while the $\Delta_R$ term tells us about the degree of correlation between the two sources of diffusion. The $\Delta_I$ term, represents a quantum drift and arises from the fact the conjugate momentum to $\varphi$ depends on $K$, and hence $\Lambda$. All in all, we see clearly how on integrating out a shell of momenta, we recreate the key structure of random walks. However, it's essential for our discussion that Equation~({\ref{eq:RG_rhoR_final}) is functional equation and not an ODE (i.e.~it is a differential equation for $\varphi(\k)$). There remains much work to translate the RG equation into the partial differential equations that describe stochastic inflation.

\subsection{Combining Time Evolution and RG}

The actual problem of interest is to integrate in a shell of comoving momentum $p \in [\Lambda,\Lambda +\Delta \Lambda]$ and evolve forward in time $a(t) \to a(t+\delta t) \approx a(t)(1 + H \delta t)$ so that the physical UV cutoff is unchanged
\beq
\Lambda_{\rm phys} = \frac{\Lambda}{a(t)} = \frac{\Lambda + \delta \Lambda}{a(t+\delta t)} \to \delta t \approx H^{-1} \frac{\delta \Lambda}{\Lambda} \ .
\eeq
The nature of this transformation is illustrated in Figure~\ref{fig:scales}. Under time evolution the reduced density matrix evolves according to 
\beq
\frac{d}{H dt} \rho_R = \frac{\partial}{H \partial t} \rho_R + \frac{\partial}{\partial s} \rho_R \ ,
\eeq
where $s = \log \Lambda$ as before. 
The first term is just the Hamiltonian evolution. Assuming our scalar field is a canonical field evolving on a potential, the Hamiltonian for the action (\ref{eq:action}) with a potential is
\beq
\hat H = \int \frac{d^3p}{(2\pi)^3} \bigg[ K \left( \frac{p^2}{\Lambda^2} \right) \bigg[\frac{1}{2 a^3} \hat \Pi^2  \bigg] + K^{-1} \left( \frac{p^2}{\Lambda^2} \right) \bigg[\frac{a^3}{2} \frac{(\nabla \vp)^2}{a^2}\bigg] + a^3 V(\vp) \bigg] + \hat H_{\rm EFT} \ , 
\eeq
where $\hat H_{\rm EFT}$ includes any higher derivative terms. The Hamiltonian is responsible for time evolution is the usual way, 
\beq
\frac{\partial}{\partial t} \rho_R = -i [ \hat H, \rho_R]  \ .
\eeq
Writing this in the $\vp_\pm$ basis, and dropping $H_{\rm EFT}$ for now, we find
\beq
\begin{aligned}
\dot{\rho}_R = \int \frac{d^3p}{(2\pi)^3} \bigg[\frac{i K}{2a^3} \left(\frac{\delta^2}{\delta \vp_+^2} - \frac{\delta^2}{\delta \vp_-^2} \right)- \frac{i a^3}{2 K}\left (\frac{(\nabla \vp_+)^2}{a^2} - \frac{(\nabla \vp_-)^2}{a^2}  \right )  - i a^3(V(\vp_+) - V(\vp_-)) \bigg]\rho_R
\end{aligned}
\eeq
Now, we can combine one step of RG and one step of time evolution to get one step of dynamical RG. 
\beq\label{eq:Master_Lindblad_eq}
\boxed{\begin{aligned}
    \frac{d}{dt} \rho_R &= \frac{H}{2}\int \frac{d^3p}{(2\pi)^3}\, \,  \bigg[\Delta_0 \left( \frac{\delta}{\delta \vp_+} + \frac{\delta}{\delta \vp_-} \right)^2 - \frac{a^6}{K^2}  \Delta_2 (\vp_- -\vp_+)^2  \\
        & + i \frac{a^3}{K} \Delta_R \bigg\{(\vp_+-\vp_-), \left(\frac{\delta}{\delta \vp_+} + \frac{\delta}{\delta \vp_-} \right) \bigg\} + 2\frac{a^3}{K}\Delta_{I} \left(  \frac{\delta}{\delta \vp_{+}}  + \frac{\delta}{\delta \vp_{-}} \right)(\vp_{+} + \vp_{-})  \bigg] \rho_R \\
        &+\int \frac{d^3p}{(2\pi)^3} \bigg[\frac{i K}{2a^3} \left(\frac{\delta^2}{\delta \vp_+^2} - \frac{\delta^2}{\delta \vp_-^2} \right)\rho_R - \frac{ia^3}{2 K}\left (\frac{(\nabla \vp_+)^2}{a^2} - \frac{(\nabla \vp_-)^2}{a^2}  \right ) \rho_R\\
        & \qquad \qquad - i a^3(V(\vp_+) - V(\vp_-))\rho_R \bigg] -i [H_{\rm EFT}, \rho_R] + \$_{\rm int} \rho_R\ .
\end{aligned}}
\eeq
For completeness, we reintroduced $\hat H_{\rm EFT}$ arising from the higher derivative contributions to $S_{\rm eff}$, and $\$_{\rm int}$ is the interaction super-operator defined in Equation~(\ref{eq:super_int}). In practice, these terms will not play a meaningful role in stochastic inflation at leading or subleading order. Therefore, we will drop $\hat H_{\rm EFT}$ and $\$_{\rm int}$ from our analysis. The remaining terms play the dominant role in the exact RG equation for the density matrix that we will analyze in the majority of the paper. 

\begin{figure}[h!]
	\centering
    \includegraphics[width=0.65\textwidth]{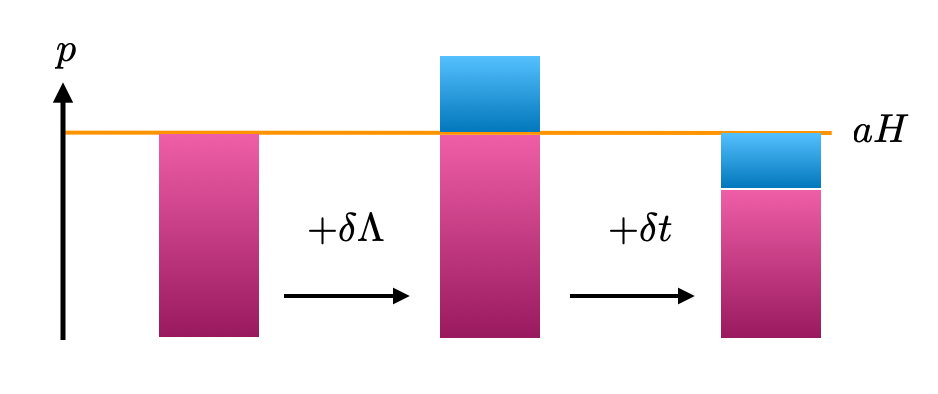}
	\caption{The combination of RG and time evolution the combined transformation holds the scale the comoving cutoff, $aH$, fixed.}
\label{fig:scales}
\end{figure}

\subsection{From Modes to Local Observables}\label{subsec:local}

One of the key issues in understanding the evolution of the density matrix is determining the operators that represent the long distance observations. To illustrate this issue and how it relates to the density matrix, let's first assume that we are in a free theory so that $S_{\rm int}=0$. Due to the boundary terms, the density matrix still evolves according to
\beq
\frac{\partial}{\partial s} \rho_R \supset\int \frac{d^3 p}{(2\pi)^3} \Delta_0(p) \left(\frac{\partial}{\partial \vp_+(\p)} + \frac{\partial}{\partial \vp_-(\p)}\right)\left(\frac{\partial}{\partial \vp_+(-\p)} + \frac{\partial}{\partial \vp_-(-\p)}\right)\rho_R[\vp_+,\vp_-] \ ,
\eeq
where $\vp_+$ and $\vp_-$ are the complete set of Fourier modes with $p,q < \Lambda(t)$. However, since $\Delta_0(p)$ only has support over a range of momenta $p\in [\Lambda, \Lambda+\Delta \Lambda]$, the RHS is a diffusion term only for modes near the cutoff. Concretely, dropping the other terms, we would get a diffusion equation of the form
\beq
\frac{\partial }{H \partial t} \rho_R= \delta(p - \Lambda(t)) \left(\frac{\partial}{\partial \vp_+(\p)} + \frac{\partial}{\partial \vp_-(\p)}\right)\left(\frac{\partial}{\partial \vp_+(-\p)} + \frac{\partial}{\partial \vp_-(-\p)}\right) \rho_R \ .
\eeq
Assuming the initial density matrix is a product between the UV and the IR modes, $\rho_0 = \rho_0(\varphi_+,\varphi_-) \otimes \rho_0(\Phi_+, \Phi_-)$, then after time $t>t_0$, we will have 
\beq
\rho_{R}(\varphi_+, \varphi_-) = {Tr}_\Phi \rho(\phi_+=\varphi_++ \Phi, \phi_- =\varphi_- + \Phi) =  \rho_0(\varphi_+,\varphi_-)  \ .
\eeq
In other words, our diffusion equation does not couple modes and therefore cannot change the state of the long wavelength modes, including generating a mixed state from a pure state. This is true of all of the quadratic boundary terms. Of course, this is a trivial statement in terms of the wavefunction for a free theory,
\beq
\Psi[\phi,t] =Z\exp\left(i\int d^3 k \frac{ (aH)^3}{2 H^2} \left(\frac{k^2}{(aH)^2} + i \frac{k^3}{(aH)^3}   \right) \phi(-\k) \phi(\k)\right)  = \prod_{k}\psi[\phi(\k)]\ .
\eeq
Here we are making the well-known observation that in a free translationally invariant theory, the Fourier modes are all independent and the wavefunction is a product state over these modes. As a result, tracing over any set of UV Fourier modes cannot create a mixed state for the remaining IR modes. 

The description in terms of stochastic inflation and the resulting mixed state for the universe, is therefore not just the result of integrating out short distance physics but also limiting our measurements to a local region. A common approach is trace over the region outside of a sphere~\cite{Maldacena:2012xp}, giving rise to non-trivial density matrix due to entanglement in space.  

In contrast, we will focus on local measurements of the long-wavelength field, $\varphi(\x)$ and its conjugate momentum $\pi(\x)$ and trace over information that leaves $\varphi(\x)$ (or $\pi(\x)$) fixed. Using translation invariance, we can choose this point to be $\x=0$ and
\bea
\varphi(\x=0) &=& \bar\varphi =\int \frac{d^3 k}{(2\pi)^3}\varphi(\k)  \\
\pi(\x=0) &=& \bar\pi =\frac{1}{N_k} \int \frac{d^3 k}{(2\pi)^3}K \left(\tfrac{k^2}{\Lambda^2} \right) \pi(\k)  \ ,
\eea
where $N_k = \int\frac{d^3k}{(2\pi)^3} K$ is the number of modes, that we introduced so that $[\bar \varphi,\bar \pi] = i$. We note that these are two local operators we can measure with a local detector. We should think of this in exact analogy with our quantum walk in Equation~\ref{eq:X_reduced}; here $\prod_\k \psi[\phi(\k)]$ is a product state of harmonic oscillators. It is only when we coupled $\sum \vp(\k)$ to an observer ($X$ in the case of the quantum walk) and trace over the harmonic oscillators that we see the mixed state emerge.

Rather than explicitly calculating the trace over $\varphi(\p)$ that leaves $\varphi(0)$ fixed, we can calculate the generating functional for $\varphi$,
\beq\label{eq:Zj_def}
Z[J_\pm,t] = {\rm Tr} \left[ e^{J_+ \bar\varphi} e^{ J_- \bar \pi} \rho_{R}(\varphi_+(\x),\varphi_-(\x))  \right] \ .
\eeq
Here it's important that we are working in the Schrödinger picture, so that all the time evolution is in the density matrix. The time evolution of this partition function is therefore given by 
\bea
\frac{d}{dt}  Z[J_\pm,t] &= {\rm Tr} \left[e^{J_+ \bar\varphi}e^{J_- \bar \pi} \frac{d}{dt} \rho_{R}(\varphi_+(\x),\varphi_-(\x))  \right]  \ .
\eea 
This representation is useful because it immediately connects many of the previous perspectives on this system. In short, by allowing $J_\pm = ik_\pm$, we can write $Z[J_\pm,t]$ as the density matrix, Wigner distribution, or other quantities associated with the variables $\bar \varphi$ and $\bar \pi$. See Appendix~\ref{app:connections} for derivations.

We can now plug in the evolution equation for the density matrix \eqref{eq:Master_Lindblad_eq}, and simplify to get the evolution equation for $Z$. The details of the calculation can be found in Appendix \ref{app:generating_func_evolution}. For the massless field, we end up with 
\begin{equation}
    \begin{aligned}\label{eq:Zevol}
        \frac{dZ[J_\pm]}{dt} &= \frac{1}{2} \int \frac{d^{3} p}{(2 \pi)^{3}}\bigg[ \left(\frac{H^{3}}{2p^{3}} J_+^2 - a^2(t_0)H \frac{p}{2} \frac{(J_-)^2}{N_k^2} - \frac{aH^2}{p} \frac{J_-J_+}{N_k} \right) \frac{\partial K}{\partial s} \bigg] Z[J_\pm]\\
        & +\frac{N_{k}}{a^{3}} J_+\left( -\frac{i J_{+}}{2} + \frac{\partial}{\partial J_{-}} \right) Z[J_\pm] - \frac{J_{-}}{N_{k}} a^3 \tilde{V}' \left(-iJ_+ \right) Z[J_\pm] \\
        &+\int \frac{d^3p}{(2\pi)^3} \left ( K\frac{a H^2}{2p} \frac{J_+ J_-}{N_k} - \frac{a^2H p}{2} K \frac{J_-^2}{N_k^2} \right)Z[J_\pm] \ .\\
    \end{aligned}
\end{equation}
While this equation still looks like a mess, it is a nice basis independent way of writing down the evolution of observables we measure in stochastic inflation. We will convert this equation to more well-known formulae in the next section.

\section{Light Fields and Stochastic Inflation}\label{sec:SI}

The description of long modes in cosmology has long been understood to mirror a random walk. In particular, as modes cross the horizon, they become classical in the sense that for a given Fourier mode, the commutator vanishes as $a \to \infty$, 
\beq
[\dot  \varphi(\k), \varphi(\vec{k}') ] = i H^2 (aH)^{-3} \delta(\k+\vec{k}') \ .
\eeq
Therefore, in an approximate sense, we must be able to treat $\varphi(\k)$ as a classical statistical variable when $k \ll aH$. This intuition can be made precise using soft de Sitter effective theory~\cite{Cohen:2020php}, where quantum mechanical effects are irrelevant by power counting.

If we start from a free theory, where each Fourier mode is independent, then $\varphi(\k)$ is like a i.i.d.~random variable. Naturally, we would expect that the variable
\beq
\bar \varphi = \varphi(\x=0) = \int \frac{d^3 k}{(2\pi)^3 } \varphi(\k) \ ,
\eeq
will behave like the location of a random walker. The goal of this section is to show precisely how this intuition arises within a fully quantum mechanical description, in analogy with the quantum walks from Section~\ref{sec:walks}.

Physically, we should interpret $\bv$ as the local measurement of the background value of the field at the point $\x=0$. For example, we could introduce a local observer with a detector whose value is measured by $\hat X$ in a pointer state~\cite{Zurek:1981xq},
\beq
|X, \varphi; t\rangle = e^{i \hat P \bv }|0\rangle \times |\Psi[\phi]\rangle \ .
\eeq
Tracing over the short distance modes then gives our random walk
\beq
\rho(X,X') = {\rm Tr} \left[ e^{ i \hat P \bv} |0\rangle\langle 0 |e^{-i \hat P \bv} \otimes\rho_R(\varphi_+, \varphi_-)  \right] \ .
\eeq
Of course, local measurements are also sensitive to short wavelength. However, we can isolate only the constant long wavelength modes using a local detector model~\cite{Green:2022fwg} that can discriminate between time dependence or wavelength. The need for the local measurement is in line with recent work that shows the need for including the observer in order to make sense of physics in dS~\cite{Chandrasekaran:2022cip,Witten:2023xze,Chen:2024rpx,Kudler-Flam:2024psh,Maldacena:2024spf,Chen:2025jqm,Blommaert:2025bgd}. This connection between the density matrix and a local observer is illustrated in Figure~\ref{fig:QW_ds}.

\begin{figure}[h!]
	\centering
    \includegraphics[width=0.65\textwidth]{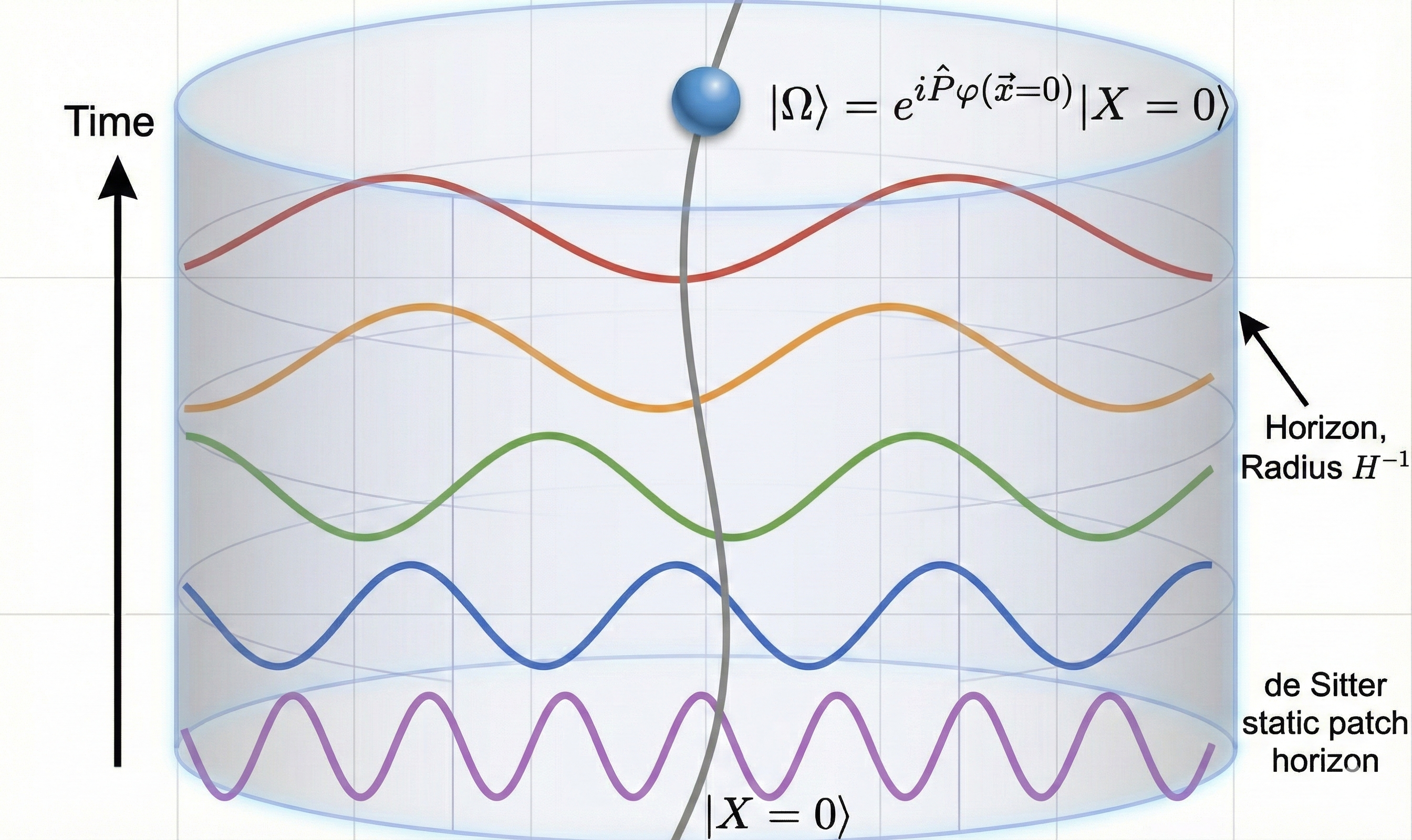}
	\caption{Visualization of a local observer measuring $\bv$ with a detector along a worldline.}
\label{fig:QW_ds}
\end{figure}

\subsection{Leading Order} 

The general strategy we will use in this section is to write the evolution equation in terms of the Wigner functional. As usual in the case of quantum random walks, the Wigner description most naturally makes contact with classical dynamical systems. 

The advantage of defining the generating functional $Z[J]$ in Equation~(\ref{eq:Zj_def}) is that it is equivalent to many representations, including the Wigner functional. Specifically, as shown in Appendix~\ref{app:connections}, we can write
\bea
W[\bar \varphi, \bar p] & = \int d \Delta \varphi e^{-i \bar p \Delta \varphi} \int \frac{dk}{(2\pi)} e^{i \frac{k\Delta \varphi}{2}} e^{-i k \bar \varphi} Z \bigg[J_+=i k, J_- = i \Delta \varphi  \bigg]  \ .
\eea
Integration by parts will allow us to convert the evolution equation for $Z[J_\pm]$ into an evolution equation for the Wigner distribution. 
\begin{equation}\label{eq:timeevZ}
    \begin{aligned}
        \frac{dZ[J_\pm]}{dt} &= \frac{1}{2} \int \frac{d^{3} p}{(2 \pi)^{3}}\bigg[ \left(\frac{H^{3}}{2p^{3}} J_+^2 - a^2(t_0)H \frac{p}{2} \frac{(J_-)^2}{N_k^2} - \frac{aH^2}{p} \frac{J_-J_+}{N_k} \right) \frac{\partial K}{\partial s} \bigg] Z[J_\pm]\\
        & +\frac{N_{k}}{a^{3}} J_+\left( -\frac{i J_{+}}{2} + \frac{\partial}{\partial J_{-}} \right) Z[J_\pm] - \frac{J_{-}}{N_{k}} a^3 \tilde{V}' \left(-iJ_+ \right) Z[J_\pm] \\
        &+\int \frac{d^3p}{(2\pi)^3} \left ( K\frac{a H^2}{2p} \frac{J_+ J_-}{N_k} - \frac{a^2H p}{2} K \frac{J_-^2}{N_k^2} \right)Z[J_\pm] \ .\\
    \end{aligned}
\end{equation}
Since the Wigner distribution is closely related to the Fourier transform of $Z[J_\pm]$, the evolution equation from $Z[J_\pm] \to W[\bar \varphi, \bar p]$ can be determined using 
\beq
J_+ = i k \to -\frac{\partial}{\partial \bar \varphi} \qquad J_- = i \Delta \varphi \to -\frac{\partial}{\partial \bar p} \qquad \frac{\partial}{\partial J_{-}} -  i \frac{J_{+}}{2} \to \bar{p} \ .
\eeq
From here, we can immediately recognize some of the familiar from describing a quantum walk (where we are rescaling $\bar p \to \bar p/N_k$)
\beq
J_+^2  \to \frac{\partial^2}{\partial \bar \varphi^2 }\qquad J_- \tilde V'(-iJ_+) \to -\frac{1}{N_k} \frac{\partial}{\partial \bar{p}} V'(\bar \varphi) \qquad J_+ \left( -\frac{i J_{+}}{2} + \frac{\partial}{\partial J_{-}} \right) \to -N_k \frac{\partial}{\partial \bar \varphi} \bar p\ .
\eeq
In contrast, the appearance of gradient terms, such as $a^{-2}(\nabla \vp)^2$, have typically not arisen in studies of stochastic inflation and corrections thereof. Intuitively, although one might imagine that gradients are small corrections, for the modes at the horizon ($p \sim aH$), this is not necessarily the case\footnote{Our analysis is most similar to the treatment of~\cite{Gorbenko:2019rza}, which similarly found additional gradient terms. However, they imposed the UV cutoff $\Lambda  = \epsilon aH$ such that these terms were suppressed by powers of $\epsilon$. This is sufficient to postpone the issue, but at the cost eliminating super-horizon modes with $\epsilon aH <p < aH$.}. Instead, as explained using SdSET, we should understand these terms as being purely scheme dependent and removable by a field redefinition. Concretely, in dimensional regularization, power-law divergences vanish, and therefore we would expect these terms to be absent as well.

The resolution here follows the same pattern as irrelevant operators in any RG. In the presence of a hard cutoff, one might generating mixing under RG between irrelevant and marginal operators, but these are removable by a change of basis. For illustration, suppose we have a scalar field theory in flat space with 
\beq
S = \int d^4 x \left( \frac{1}{2} \partial_\mu \phi \partial^\mu \phi - \frac{\lambda_4}{4!}\phi^4 - \frac{\lambda_6}{6!} \frac{1}{\Lambda^2}\phi^6 \right) \ .
\eeq
With a hard cutoff $\epsilon\Lambda$, one a one-loop contribution to $\lambda_4$ $\beta$-function
\beq
\frac{\partial}{\partial \log \Lambda} \lambda_4 \supset \frac{\lambda_6}{4\pi^2} \epsilon^2 \ .
\eeq
Here we have introduced $\epsilon$ to show how it depends on the value of the hard cutoff, but ultimately we should take $\epsilon \to 1$ as $\Lambda$ is defined to be our RG scale. The key to eliminating this term is to realize that $\lambda_6$ has a classical $\beta$-function so that
\beq
\frac{\partial}{\partial \log \Lambda} \lambda_6 = 2 \lambda_6 \to \frac{\partial}{\partial \log \Lambda} (\lambda_4 - \frac{\lambda_6}{8\pi^2} \epsilon^2 ) \equiv \frac{\partial}{\partial \log \Lambda} \tilde \lambda_4  = 0  \ . 
\eeq
We see that by redefining the coupling as $\tilde \lambda_4$, we can eliminate the apparent contribution to RG from $\lambda_6$. Of course, when we regulate with dimensional regularization, the power-law divergence vanishes, and there is no need to redefine the coupling.

The evolution of the density matrix under dynamical RG is in precise analogy with the scalar field example. Using $N_k \approx (aH)^3$ so that $\dot N_k = 3H N_k$, we notice that 
\beq
\begin{aligned}
\frac{1}{H}\frac{\partial}{\partial t}\left[\left(- a^2(t_0)H \frac{p}{2} \frac{(J_-)^2}{N_k^2} - \frac{aH^2}{p} \frac{J_-J_+}{N_k} \right) + \left ( K\frac{a H^2}{2p} \frac{J_+ J_-}{N_k} - \frac{a^2H p}{2} K \frac{J_-^2}{N_k^2} \right) \right] = \\
\left[\left( 4 a^2(t_0)H \frac{p}{2} \frac{(J_-)^2}{N_k^2} +2 \frac{aH^2}{p} \frac{J_-J_+}{N_k} \right) + \left ( -2 K\frac{a H^2}{2p} \frac{J_+ J_-}{N_k} + 4 \frac{a^2H p}{2} K \frac{J_-^2}{N_k^2} \right)\right] \ . \label{eq:tdep_terms}
\end{aligned}
\eeq
In contrast, the additional terms in Equation~(\ref{eq:timeevZ}) do not explicitly depend on time. We can therefore remove the terms in Equation~(\ref{eq:tdep_terms}) entirely by a polynomial redefinition of $Z[J_\pm]$.

Removing the irrelevant\footnote{Here we mean irrelevant in the technical RG sense, as follows from SdSET.} terms in the equation by this redefinition, we are left with the evolution equation for the Wigner distribution 
\begin{equation}
    \begin{aligned}
        \frac{d}{dt}W[\bar \varphi, \bar p] = \frac{H^3}{8 \pi^2} \frac{\partial^2}{\partial \bar \varphi^2} W[\bar \varphi, \bar p]- \frac{\bar p}{a^3} \frac{\partial}{\partial \bar \varphi} W + a^3 \frac{\partial}{\partial \bar p}V'(\bar \varphi)W \ .
    \end{aligned}
\end{equation}
It is instructive to compare the final Wigner evolution equation to the density matrix evolution that we derived from QFT, and see which terms survived. We started with density matrix evolution in \eqref{eq:Master_Lindblad_eq}, and see that the ultimate fate is that for local measurements, the following terms drop out due to being irrelevant
\begin{align}
    \frac{d}{dt} \rho_R &= \frac{H}{2}\int \frac{d^3p}{(2\pi)^3}\, \,  \bigg[\Delta_0 \left( \frac{\delta}{\delta \varphi_+} + \frac{\delta}{\delta \varphi_-} \right)^2 - \textcolor{gray}{ \frac{a^6}{K^2} \Delta_2 (\varphi_- -\varphi_+)^2}  \\
        & +  \textcolor{gray}{i \frac{a^3}{K} \Delta_R \bigg\{(\varphi_+-\varphi_-), \left(\frac{\delta}{\delta \varphi_+} + \frac{\delta}{\delta \varphi_-} \right) \bigg\}} + 2 \frac{a^3}{K}\Delta_{I} \left(  \frac{\delta}{\delta \varphi_{+}}  + \frac{\delta}{\delta \varphi_{-}} \right)(\varphi_{+} + \varphi_{-})  \bigg] \rho_R \nonumber \\
        &+\int \frac{d^3p}{(2\pi)^3} \bigg[\frac{i K}{2a^3} \left(\frac{\delta^2}{\delta \varphi_+^2} -  \frac{\delta^2}{\delta \varphi_-^2} \right)\rho_R - \ \textcolor{gray}{\frac{ia^3}{2 K}\left (\frac{(\nabla \varphi_+)^2}{a^2} - \frac{(\nabla \varphi_-)^2}{a^2}  \right ) \rho_R} - i a^3(V(\varphi_+) - V(\varphi_-))\rho_R \bigg] \nonumber \ .
\end{align}
Out of the remaining terms, we see 
\begin{equation}
    \begin{gathered}
        \int \frac{d^3p}{(2\pi)^3}\bigg[ \Delta_0 \left( \frac{\delta}{\delta \varphi_+} + \frac{\delta}{\delta \varphi_-} \right)^2 + 2 \frac{a^3}{K} \Delta_{I} \left(  \frac{\delta}{\delta \varphi_{+}}  + \frac{\delta}{\delta \varphi_{-}} \right)(\varphi_{+} + \varphi_{-}) \bigg] \rho_R  \to\frac{H^3}{8 \pi^2} \frac{\partial^2 W}{\partial \bar \varphi^2} \\
        \int \frac{d^3p}{(2\pi)^3} \bigg[\frac{i K}{2a^3} \left(\frac{\delta^2}{\delta \varphi_+^2} -  \frac{\delta^2}{\delta \varphi_-^2} \right)\rho_R  \bigg] \to - q \frac{\partial W}{\partial \bar \varphi} \\
        -\int \frac{d^3p}{(2\pi)^3} i a^3(V(\varphi_+) - V(\varphi_-))\rho_R  \to a^3 \frac{\partial}{\partial \bar p}V'(\bar \varphi)W  \ .
    \end{gathered}
\end{equation}
It's useful to remember that $\Delta_I$ contains the boundary term that determines the real part of the wavefunction, as shown around Equation~(\ref{eq:total deriv}). As a result, it is directly related to the Gaussian fluctuations of each mode and thus will be crucial in the derivation of the Fokker-Planck equation. See Appendix \ref{app:generating_func_evolution} for details of this derivation.

Now, we can transform variables to $\bar p = a^3(t) q$ to get
\begin{equation}
    \frac{d}{dt}W[\bar \varphi, q] = \frac{H^3}{8 \pi^2} \frac{\partial^2 W}{\partial \bar \varphi^2} + 3H \frac{\partial}{\partial q}(q W) - q \frac{\partial}{\partial \bar \varphi} W +  \frac{\partial}{\partial q}V'(\bar \varphi) W\ .
\end{equation}
Having set up the full phase space evolution equation, we can now compare this to the walk introduced in Section \ref{sec:walks}. We see that this reproduces the same exact differential equation with $D = H^3/8 \pi^2$, $\gamma = 3H, m = 1$\footnote{We see that stochastic inflation is represented as a noisy $x$ and deterministic $p$ evolution. This agrees with previous results for the evolution of the density matrix~\cite{Nambu:1991vs,Li:2025azq}, but contradicts some generalizations of Stochastic Inflation~\cite{Woodard:2005cw,Tolley:2008na,Moss:2016uix}. Generalizations exist, where even $p$ can be noisy due to coupling to additional degrees of freedom. These are studied in the Open EFT of Inflation. See for instance \cite{Salcedo:2024smn} and references therein.}. As explained, at leading order, we can integrate out $q$ by setting $q = -V'(x)/3H$. We then get the well known stochastic inflation result 
\begin{equation}\label{eq:SI_class}
    \frac{dP}{dt} = \frac{H^3}{8 \pi^2} \frac{\partial^2 P}{\partial \bv^2} + \frac{\partial}{\partial \bv} \left( \frac{V'(\bv)}{3H} P(\bv) \right) \ .
\end{equation}
There are, of course, easier ways to arrive at this result. What is meaningful here is that the determination of this PDE from the exact RG as a functional over all the Fourier modes is more nuanced than one might have expected from just comparing Equations~(\ref{eq:final_RG_quad}) and~(\ref{eq:compare_RW}). 

\subsection{Next-to-Leading Order}

Our goal in this section is to determine the leading corrections to the stochastic formalism. However, unlike perturbation theory around the vacuum, the fluctuations around a random walk have a non-trivial definition of leading order (LO), next to leading order (NLO) and so on, that requires explanation.

Stochastic inflation is conventionally defined in terms of $P(\bar \varphi,t)$, the probability distribution for the long wavelength field. In terms of the Wigner distribution, we get this probability by ``marginalizing" over $q$:
\beq
P(\bar \varphi,t) =\int dq \, W(\bar \varphi, q, t) \ .
\eeq
If we start with our LO description of stochastic inflation, equilibrium probability distribution can be determined by demanding $\dot P= 0$ and yields
\begin{align}
P_{\rm eq}(\bar \varphi) = \int dq W_{\rm eq}(\bar \varphi, q) &= \tilde A \exp\left(-\frac{8\pi^2 V(\bar \varphi)}{3H^4} \right) \ .
\end{align}
Let's take the specific case where $V(\bar \varphi) = \frac{\lambda}{4!}\bar \varphi^4$. The equilibrium probability distribution allows for order one probabilities for any $|\bar\varphi| < \lambda^{-1/4} H$. One might be concerned that there is no reliable expansion in $\lambda \ll 1$ given that $\bar \varphi  \to \infty$ as $\lambda \to 0$. Fortunately, the corrections come with positive powers of $\lambda$ that a meaningful expansion exists. At leading order, the time evolution of the probability distribution, Equation~(\ref{eq:SI_class}), scales as $\frac{d}{dt} P = {\cal O}(\lambda^{1/2})$ when $\varphi \sim \lambda^{-1}$. At next-to-leading order (NLO), the correction to the time evolution gives contributions to $\dot P$ that are ${\cal O}(\lambda)$, and are therefore suppressed by $\sqrt{\lambda} \ll 1$ as desired.

Previous work has calculated the NLO corrections to the Stochastic formalism and to the evolution of the density matrix. At this order, corrections to the Fokker-Planck equation can be defined as a non-trivial\footnote{Here we are avoiding calling these corrections universal because they are sensitive to the split between the time-independent and time-dependent components of $\varphi$. Specifically, the correction below is correct for the definition used in~\cite{Mirbabayi:2020vyt,Cohen:2021fzf} but appears to contradict~\cite{Gorbenko:2019rza}. The difference arises from a redefinition of the decaying mode and can be accounted for by matching~\cite{Cohen:2021fzf}. } correction to the effective potential~\cite{Gorbenko:2019rza,Mirbabayi:2020vyt,Cohen:2021fzf}
\beq
V_{\rm eff, NLO}(\bv) = \frac{\lambda}{4!} \bv^4 + \frac{1}{6} \frac{1}{3!}\frac{\lambda^2}{18 H^2} \bv^6 + {\cal O}(\lambda^3 \bv^8) \ .
\eeq
Notice that the effective potential becomes a series expansion in $\phi^2 \lambda$ and therefore breaks down when $\phi \sim 1/\sqrt{\lambda}$. The effective potential corrections at NLO and beyond turn out to be determined by a field redefinition in SdSET. Concretely, eliminating the decaying mode, which is the conjugate momentum, introduces these terms. It was argued in~\cite{Cohen:2022clv} that the breakdown of series expansion of $V_{\rm eff}$ arises because the conjugate momentum to $\phi$ cannot be ignored at large field values.

The role of the momentum in generating the effective potential is seen most directly if we interpret the evolution of the Wigner distribution as a Langevin system. This is in exact analogy to the walk we analyzed in \ref{subsec:model_si}. In that case, $\dot {\bar\varphi }= q + \xi$ where $\xi$ is our noise and 
\beq
\dot q = 3 H q + V'(\bar \varphi) \to q(t) = \int^t dt' V'(\bar \varphi(t'))e^{3H(t-t')} \ .
\eeq
Now expanding $\varphi(t') = \varphi(t) + (t'-t) \dot \varphi(t) + \ldots$ we have
\beq
q(t) = - \frac{V'(\bar \varphi)}{3H} + \dot \phi V''(\bar \varphi) \int dt (t'-t) e^{3H(t-t')}  = - \frac{V'(\bar \varphi)}{3H} - \frac{1}{27 H^3} V'(\bar \varphi)V''(\bar \varphi) \ .
\eeq
For $V(\bar\varphi) = \frac{\lambda \bv^4}{4!}$, if we interpret $3 H q = - V_{\rm eff}'$, then we can see that 
\beq
V_{\rm eff} =\frac{\lambda}{4!} \bv^4 +\frac{1}{6}  \frac{1}{3!}\frac{\lambda^2}{18 H^2} \bv^6  + \ldots \ ,
\eeq
in exact agreement with the direct calculations in QFT.

Of course, we should be able to derive a correction to the Fokker-Planck equation from $q$ without the need for a Langevin description,
\begin{equation}
    \frac{d}{dt}W[\bar \varphi, q] = \frac{H^3}{8 \pi^2} \frac{\partial^2 W}{\partial \bar \varphi^2} + 3H \frac{\partial}{\partial q}(q W) - q \frac{\partial}{\partial \bar \varphi} W +  \frac{\partial}{\partial q}V'(\bar \varphi) W\ .
\end{equation}
However, since the evolution of $q$ is purely deterministic at this order, we can use the intuition from the Langevin description to define a change of variables
\beq
Q =a^3(t) \left(q +\frac{1}{3H} V'_{\rm eff}(\bar \varphi) \right)
\eeq
so that 
\bea
\frac{\partial}{\partial t }&\to& \frac{\partial Q}{\partial t} \frac{\partial}{\partial Q } + \frac{\partial}{\partial t } =  3 H Q \frac{\partial}{\partial Q} + \frac{\partial}{\partial t} \\
\frac{\partial}{\partial \bar \varphi }&\to& \frac{\partial Q}{\partial \bar\varphi} \frac{\partial}{\partial Q } + \frac{\partial}{\partial \bar \varphi } = \frac{a^3}{3H} V_{\rm eff}''(\bar \varphi) \frac{\partial}{\partial Q }+ \frac{\partial}{\partial \bar \varphi } \\
\frac{\partial}{\partial q } &\to & \frac{\partial Q}{\partial q }\frac{\partial}{\partial Q} =a^3 \frac{\partial}{\partial Q } \ .
\eea
Making this substitution, we find
\begin{align} 
\frac{d}{dt} W[\bar\varphi, q] &= 3HW + \frac{H^{3}}{8\pi ^{2}} \frac{\partial^2 W}{\partial \bv ^{2}} +  a^{3} V_{\mathrm{eff}}'' \frac{ H^{2}}{12 \pi ^{2}}  \frac{\partial^2 W}{\partial Q \partial \bv }  + a^6 \frac{H}{72 \pi ^{2}} (V_{\mathrm{eff}}'')^{2} \frac{\partial ^{2} W}{\partial Q^{2}} \nonumber\\
&  + a^{3}\frac{\partial}{\partial Q} \left(- V_{\mathrm{eff}}' + \frac{1}{9H^{2}}V_{\mathrm{eff}}' V_{\mathrm{eff}}'' +  \frac{ H^{2}}{24 \pi ^{2}} V_{\mathrm{eff}}'''  + V' \right) W \nonumber\\
& -  V_{\mathrm{eff}}'' \frac{\partial}{\partial Q} \left(\frac{Q}{3H} W \right) - \frac{Q}{a^{3}} \frac{\partial W}{\partial \bv} + \frac{1}{3H}  \frac{\partial}{\partial \bv} (V_{\mathrm{eff}}' W)
\end{align} 
In order to have a normalized distribution in these new variables, we now rescale the Wigner distribution
\beq
\tilde W[\bar \varphi, Q] =\frac{\partial q}{\partial Q}W[\bar \varphi, q] = a^{3}(t) W[\bar \varphi, q]
\eeq
We can remove the mean for $Q$ by choosing $V_{\rm eff}$ (neglecting $V_{\rm eff}'''$ at the order we want to consider) so that
\beq
V'_{\rm eff}(1- \frac{1}{9H^2} V_{\rm eff}'') = V' \to V_{\rm eff}'(\bar \varphi) \approx V'(\bar \varphi) + \frac{1}{9H^2} V'(\bar \varphi)V''(\bar \varphi) \ .
\eeq
The final step to determining the evolution of the probability of $\bv$ is expand in moments~\cite{Li:2025azq}, defined by
\beq
 W_N(\bv) = \int dQ \left( \frac{Q}{a^3}\right)^N \tilde W(\bv,Q) \qquad P(\bv) \equiv W_0(\bv) \ .
\eeq
This is effectively the same procedure used to convert the Boltzmann equations to the fluid equation, as is commonly done for describing the CMB\footnote{See~\cite{Pan:2016zla}, for example, for a systematic expansion of the corrections to the tight coupling approximation similar to our approach.}. After performing our change of variables, this procedure is straightforward and gives rise to the NLO evolution equation
\begin{equation}\label{eq:NLO_final}
    \frac{dP}{dt} = \frac{H^3}{8 \pi^2} \frac{\partial^2 P}{\partial \bar \varphi^2} + \frac{\partial}{\partial \bar \varphi} \left( \frac{V_{\rm eff}'(\bar \varphi)}{3H} P(\bar \varphi) \right) + \frac{1}{3H}\frac{\partial}{\partial \bv} \left( V''_{\rm eff} \frac{H^2}{12 \pi^2}\frac{\partial}{\partial \bv} P(\bv) \right)
\end{equation}
The final term arises from solving for the evolution of $W_1$ at NLO and plugging it into the equation for $W_0$. As a check, the equation agrees with the equivalent equations derived in Section~\ref{subsec:model_si}. The details of this calculation are shown in Appendix~\ref{app:moments}. In the $\lambda \bv^4$ theory, the final term is removable by a field redefinition~\cite{Cohen:2021fzf} of $\bv$ and thus is not universal.

The essential take-away from this calculation is that all of the information relevant to both the LO and NLO calculation of stochastic inflation is present in the master equation for the Wigner distribution without including higher-order corrections of $S_{\rm eff}$ in our exact Wilsonian scheme.

\subsection{NNLO and Beyond}

When calculating the corrections to Stochastic Inflation from QFT, the first non-trivial correction arises at NNLO, which is~${\cal O}(\lambda^{3/2})$ in the Fokker-Planck equation. Specifically, at this order we generate a non-Gaussian noise term
\begin{equation}
    \frac{dP}{dt} = \frac{H^3}{8 \pi^2} \frac{\partial^2 P}{\partial \bar \varphi^2} + \frac{\partial}{\partial \bar \varphi} \left( \frac{V_{\rm eff}'(\bar \varphi)}{3H} P(\bar \varphi) \right) + \frac{d_0}{3!}\frac{\partial^3}{\partial \bar \varphi^3}\left(\bar \varphi  P( \bar \varphi) \right) \ ,
\end{equation}
where $d_0= \frac{\lambda}{192\pi^2} $ was calculated in~\cite{Cohen:2021fzf}. However, this result has not been obtained in a fully consistent scheme that removes all the sub-divergences. See~\cite{Beneke:2023wmt} for progress in this direction.

An essential difference between the approach in this paper and~\cite{Cohen:2021fzf} is that we are working in a Wilsonian EFT, rather than a continuum EFT~\cite{Georgi:1993mps}. In the Wilsonian description, we integrated out the short modes explicitly, giving rise to higher dimension interactions, including higher derivative terms, in $S_{\rm eff}$. These terms do not arise in a continuum theory, as they would vanish in dimensional regularization. Nevertheless, the corrections to these terms in the Wilsonian description are needed to reproduce higher order contributions to the RG flow~\cite{Polchinski:1983gv}. In this regard, the challenge of calculating NNLO corrections in our exact RG scheme is not unique to de Sitter, but a common aspect of Wilsonian versus continuum descriptions.

In addition to these higher order terms from $S_{\rm eff}$, we expect additional terms from eliminating $Q$ systematically. These terms are also generally scheme dependent and will bring higher powers of $\lambda$. There is a unique contribution at NNLO that is both straightforward to calculate, and could contribute to $d_0$, that arises from
\beq
\frac{d}{dt}\tilde W(\bv,Q)\supset - \frac{1}{24} V'''(\bv) \frac{\partial^3}{\partial Q^3} \tilde W \ .
\eeq
Solving for the higher moments of $\tilde W$ and reducing the equation to the solution of $P = W_0$, one finds the NNLO term
\beq
\frac{d}{dt} P  \supset \frac{1}{648 H^3} \frac{\partial^3}{\partial \bv^3}\left( V'''(\bv) P \right)
\eeq
See Appendix~\ref{app:moments} for the derivation. Notice that using $V = \lambda \bv^4/4!$, this is a contribution to $d_0$ defined above but is not consistent with the two-loop calculation in~\cite{Cohen:2021fzf}. Moreover, at NNNLO, we expect a contribution from a three loop diagram that would give a four derivative correction proportional to $\lambda$. There is no such term that can arise from eliminating $Q$.

The fact that the higher order corrections appear to require a more detailed analysis of the RG evolution is consistent with the physical expectation that higher-derivatives terms arise from non-Gaussian noise. Specifically, in a Markovian random walk, these terms arise from non-Gaussian cumulants in the Kramers-Moyal expansion of the transition amplitudes~\cite{gardiner2004handbook}. It would therefore be surprising to fully capture this effect, which we associate with the behavior of modes at horizon crossing, purely from the elimination of the fast variables in the super-horizon description~. 

Technically speaking, SdSET offers a much more straightforward path to calculating the coefficients at NNLO and beyond. However the long term goal of this program is to calculate the tail of the distribution of $\bv$ beyond the limits of SdSET, when $\lambda \bv^2 > 1$. This will likely\footnote{An alternate proposal for calculating the tail of the distribution was made in~\cite{Celoria:2021vjw} directly in terms of the wavefunction. As the statistics of local operators are described by a density matrix, an effective description in terms of a pure state can miss essential contributions to these observables~\cite{Green:2024cmx}.} require a hybrid approach of SdSET and exact RG. The tails are sensitive to UV physics and therefore cannot be calculated purely within an effective description~\cite{Cohen:2021jbo,Cohen:2022clv} (see also~\cite{Panagopoulos:2020sxp,Achucarro:2021pdh} for specific models). Yet, for a theory like $\lambda \phi^4$, where the theory is renormalizable, one might expect to be able to describe the tail up to the Landau pole. This is reminiscent of the effective potential in flat space, which can be improved using RG to resum the log corrections. There is reason to suspect a similar approach could apply for large deviations in de Sitter, but we will leave this to future work. 

\section{Renomalization in dS} \label{sec:RG}

Although light fields are where the need for normalization is most acute, our understanding of QFT in dS requires RG for all masses and spins, just like flat space. Direct calculations of the spectral densities are naturally interpreted in the usual language of conformal invariance and anomalous dimensions~\cite{Hogervorst:2021uvp,DiPietro:2021sjt,Chakraborty:2023qbp,Loparco:2023rug}. Much like the case of physics in AdS, it is natural to expect the language of RG is useful for interpreting these results.

One strong piece of evidence that RG flow can be used directly in dS comes from conformal field theory: if a given theory is conformal in flat space,  then the same theory in dS will exhibit the same list operators and scaling dimensions. This can be seen easily using Weyl transformation. However, at non-trivial fixed point, these dimensions arise from mixing between operators and/or anomalous dimensions generated by interactions and therefore the same RG resummation needed to describe the CFT in flat space must apply in dS as well~\cite{Green:2020txs}. 

In flat slicing of dS, the Weyl transformation that takes flat space correlators to cosmological correlators involves scaling by the scale factor $a(\tau)$. For a CFT, this requires that any anomalous dimensions are associated with time-dependent logs. The required matching between dS and flat space therefore strongly suggests that the RG resummation involves time-dependent logs in dS. In this sense, it is dynamical RG that should be on the same footing as conventional RG in flat space\footnote{For principle series fields, some anomalous dimensions can be calculated by resumming bubble diagrams, more like a decay width~\cite{Marolf:2010zp,Chakraborty:2023eoq}. These examples are compatible with the RG description, as shown in~\cite{Cohen:2024anu,Chakraborty:2025mhh}. Yet, the conformal example~\cite{Green:2020txs} demonstrates that not all anomalous dimensions can be calculated by summing geometric series.}.

In this section, we will demonstrate that dynamical RG in dS obeys these properties. We will show that the exact RG argument of Section~\ref{sec:exact} can be applied to massive fields without any meaningful changes to the evolution equation. Repeating the argument of Polchinski~\cite{Polchinski:1983gv} then demonstrates that all the RG corrections can be absorbed into the definitions of couplings and local operators of the theory.

\subsection{Scalar Fields of Any Mass}

Our starting point will be an interacting scalar field $\varphi(\x)$ with a mass $m$ such that the action and density matrix are defined by Equation~(\ref{eq:action}), now with
\beq
\begin{aligned}
S_0[\varphi(\vec{x}, t), \Lambda] & = \int_\gamma d t a^3(t) \int \frac{d^3 k}{(2 \pi)^3} \frac{1}{2} K^{-1}\left(\frac{k^2}{\Lambda^2}\right)\left(\nabla_\mu \varphi(\k,t)  \nabla^\mu \varphi(\x,t)- m^2 |\varphi(\k,t)|^2 \right) \ .
\end{aligned}
\eeq
We be be concerned only with local operators constructed from $\varphi(\x)$ and derivatives, ${\cal O}_n(\x)$. Since we are interested in local correlation function in the in-in formalism (rather than the density matrix, following~\cite{Green:2024cmx}, we will define a partition function
\beq
Z[\{ J_n \};t] ={\rm Tr}_\phi \left[ \exp\left(\sum_n \int d^3 x J_n {\cal O}_n(\x)\right) \, \rho(\phi,\phi')  \right] \ ,
\eeq
where $\O_n(\x)$ are local operators composed from the UV field $\phi$ in the Schrodinger representation. Because of the trace, we know the result must be independent of $\Lambda$, in contrast to the density matrix itself. Our goal is to show that the process of integrating out the short distance physics and evolving in time combine to give a local theory where the local action and operators are modified. 

At this stage, we can repeat the argument from Section~\ref{sec:exact} with only minor modifications. Nothing about the procedure assumed the fields were massless, other than the specific form of the action and Green's function. However, we note that if we make the substitutions
\beq
\nabla^2 \to \nabla^2 - m^2 \qquad G^{-1} = K^{-1} (\nabla^2 -m^2) \qquad G^{-1} G = a^{-3}\delta(t-t')\delta(\x-\x') \ ,
\eeq
then exact RG result is unchanged. The essential reason is that the mass term, or any potential $V(\phi)$, does not introduce any boundary terms when integrating by parts. The boundary terms are again defined by $\Delta_{0}$, $\Delta_2$, and $\Delta_{\rm I, R}$ with the Green's functions replaced by the massive ones, for example
\bea
 \Delta_0(p,t_0) &=&iG' = iG'(t_0,t_0,\p) =  -\frac{\partial K}{\partial s} \frac{\pi}{4} (-\tau_0)^3 H^{(1)}_{\nu}(-p\tau_0) H^{(2)}_{\nu}(-p\tau_0) \ ,
\eea
where $\tau_0$ is the conformal time at physical time $t_0$.

\subsection{Diagrammatic understanding of Exact RG}

Given the setup, we can now apply exact RG to the dS generating functional directly. Our goal is to demonstrate that the change to $Z[\{ J_n \};t]$  from integrating out $p \in [\Lambda, \Lambda +\delta \Lambda]$ and shifting in time by $\delta t = H^{-1}\delta \Lambda / \Lambda$ can be completely absorbed into the definition of (a) the effective action $S_{\rm eff}$ and (b) the definition of the local operators $\O_n(\x)$. This will ensure that our RG equations are differential equations with no explicit dependence on the UV physics and therefore can be integrated to determine correlation functions, just like the Callan–Symanzik equation for RG in flat space.

The first step is straightforward: we have already shown that integrating out a shell of momenta gives rise to the following change to the effective action, illustrated in Figure~\ref{fig:exactRG},
\beq
\frac{\partial}{\partial s} S_{\rm int}= \int \frac{d^3 p}{(2 \pi)^3} \int_\gamma d t d t^{\prime}\left(\frac{1}{2} \frac{\delta S_{\rm int}}{\delta \varphi(t)} G\left(t, t^{\prime}\right) \frac{\delta S_{\rm int}}{\delta \varphi\left(t^{\prime}\right)}-\frac{1}{2}\left(\frac{\delta^2 S_{\rm int}}{\delta \varphi(t) \delta \varphi\left(t^{\prime}\right)} G\left(t, t^{\prime}\right)\right)\right) \ . 
\eeq
The second term is manifestly local because
\beq
\frac{\delta^2 S_{\rm int}}{\delta \varphi(t) \delta \varphi\left(t^{\prime}\right)} = \delta(t-t^{\prime}) \frac{\partial^2}{ \partial^2 \varphi} {\cal L}_{\rm int}(\varphi,t) \ .
\eeq
The first-term, on the other hand, is bi-local in time and therefore technically non-local. However, with the $i\epsilon$ prescription, we know that 
\beq
G(t,t')\propto e^{-\epsilon p|t-t'|} \ ,
\eeq
for $t,t'$ on the same side of the Schwinger-Keldysh contour (i.e. $t,t' \in (-\infty(1-i\epsilon),t_0]$ or $t,t \in (-\infty(1+i\epsilon),t_0]$)
This ensures we can Taylor expand in $t-t'$ and integrate each term to find a finite result to all orders. In this sense, as far as the effective action is concerned, there is nothing about integrating out a shell of momentum in dS that is distinct from flat space.

The situation is similar from the contributions to the action from $t \in (-\infty(1-i\epsilon),t_0]$ and $t'  \in (-\infty(1+i\epsilon),t_0]$ (or vise versa). In these cases, the Green's functions are again exponentially suppressed but now by 
\beq
G(t=t_+,t'=t_-)\propto e^{-\epsilon p|t_+-t_0| - \epsilon p |t_- - t_0| } \ .
\eeq
These terms should therefore be expanded around $t_\pm = t_0$, giving rise to a local expansion in operators at the boundary. These terms are responsible for renormalizing the definition of local operators, or equivalently, the sources $J_{n}$.

The non-trivial changes that arise for in-in correlators in de Sitter space are the boundary terms at $t_0$ and the fact that $\Lambda(t)$ is combined with time-evolution in the full RG transformation. Yet, it is straightforward to understand the impact of these terms. Writing our generating functional as 
\beq
\frac{d}{dt} Z[\{ J_n \};t] = {\rm Tr}_\varphi \left[ \exp\left(\sum_n \int d^3 x J_n {\cal O}_n(\x)\right) \, \frac{d}{dt} \rho_R(\varphi,\varphi')  \right] \ .
\eeq
Of course, this is identical to the derivation of the evolution in Equation~(\ref{eq:Zevol}), so that schematically we have
\beq
\frac{d}{dt}  Z[J_\pm,t] =  {\rm Tr} \left[\exp\left(\sum_n \int d^3 x J_n {\cal O}_n(\x)\right) \left( - i [H_{\rm eff}, \rho] + \sum_i L_i \rho L^\dagger_i  - \frac{1}{2} \{L_i^\dagger L_i,\rho)\}  \right)  \right]  \ .
\eeq
Using the trace, we can exchange operators acting on $\rho$ to acting on the local operator insertions. However, because these act locally on ${\cal O}_n(\x)$ we can express time evolution as a local change to the operator (which would be the Heisenberg picture, in the absence of the $L_i$). Therefore, we can treat this apparent time evolution as the operator equation for
\bea
\frac{d}{dt} V[\{J_n\},t] &=& i [H_{\rm eff} ,V[\{J_n\},t] ] + \sum_i L^\dagger_i  V[\{J_n\},t] L_i - \frac{1}{2}  \{L_i^\dagger L_i,V[\{J_n\},t]\}  \\
V[\{J_n\},t] &=& \exp\left(\sum_n \int d^3 x J_n {\cal O}_n(\x)\right) \ .
\eea
If ${\cal O}_n$ forms a complete basis of local operators, then the change to $V[\{J_n\},t]$ is just a redefinition of $J_n$. This expression is purely local in time and therefore we can simply redefine $J_n \to J_n'$ to capture the time evolution of the system.

One might worry that the fact that $J_n(\x)$ is a source that depends directly on position could introduce non-locality in space. This does not occur because the non-trivial evolution of the operators arises from equal time commutators, and therefore are local in space by causality. This is manifest in the UV Hamiltonian description. In terms of the Wilsonian RG, the evolution of the operators arises from contribution that couple the two branches of the Schwinger-Kelgysh contour ($t_+$ and $t_-$) and are not equivalent to a local term in the action because of the path ordering of the exponential. Of course, the time ordering is only relevant for the terms that don't commute, which are again local in space at equal time. The same features appear in flat space where several examples have been worked out in detail~\cite{Green:2024cmx}.

The essential result here again follows from Equation~(\ref{eq:exactRG_def}). The sources can be added to the boundary action
\beq
S_{\partial, {\rm tot}}= S_\partial + \int d^3 x \sum_n J_n\O_n(\x) \ .
\eeq
In these variables, our partition function is simply 
\beq
Z[\{ J_n \};t]  = \int_\gamma D \varphi T \exp \left(i S_0+i S_{\rm int}  + i S_\partial \right)  \ ,
\eeq
and therefore obeys that same renormalization properties as the effective action. At this stage, the properties of the renormalization of de Sitter correlators and the flat space in-in correlators are identical, up to the definition of Green's functions and the fact that the de Sitter RG transformation includes a step of time evolution.

It is, perhaps, worth noting that the analogous proof of ``renormalizability" in AdS remains incomplete~\cite{Heemskerk:2010hk}. In that case, we have the benefit of AdS/CFT which implies that the dual description is a conventional QFT and therefore the bulk of AdS should inherit what is known renormalization. Yet, some of the peculiar aspects of dynamical RG in dS~\cite{Chakraborty:2025mhh} have nearly identical avatars in AdS~\cite{Fitzpatrick:2012cg,Baumann:2019ghk}. A proof in AdS would likely be instructive both for understanding holographic RG in AdS~\cite{Skenderis:2002wp}, and as another connection between physics in AdS and dS.

\section{Conclusions}\label{sec:conclusions}

Understanding the dynamics of light scalar fields on a fixed de Sitter background presents a surprising challenge. Perturbative calculations fail due to large time-dependent logs~\cite{Akhmedov:2019cfd,Green:2022ovz}. These logs have been understood and resummed using the ideas from classical random walks, but leave the quantum nature of the fields obscure. Moreover, while understanding light scalars should be a stepping stone to quantum cosmology, the stochastic formalism bears little resemblance to the Euclidean path integral or wavefunction of the universe~\cite{Hartle:1983ai}. Yet, both approaches yield the same equilibrium probability distribution~\cite{Maldacena:2024uhs}, which demands a deeper explanation. A comparison of the density matrices of the no-boundary proposal~\cite{Page:1986vw,Hawking:1986vj,Fumagalli:2024msi,Ivo:2024ill} with the stochastic formalism offers the opportunity to make a deeper connection between these descriptions.

Understanding the de Sitter entropy~\cite{Gibbons:1977mu}, entanglement of inflationary modes,~\cite{Maldacena:2012xp,Abate:2024xyb,Boutivas:2024lts}, and cosmology as an open quantum system~\cite{Burgess:2015ajz,Salcedo:2024smn} motivates the need to describe the evolution of the density matrix from a stochastic perspective. In this paper, we developed the tools of exact RG to provide a non-perturbative evolution equation for the long-wavelength density matrix. While elements of the stochastic description arise from tracing out sub-horizon fluctuations, we also find that the description as a quantum walk only emerges when additionally tracing over information at separated points. This formalism clarifies and generalizes various aspects of stochastic inflation and should serve a powerful tool for exploring non-perturbative aspects of QFT in dS.

The density matrix in dS also played a significant role in prior work~\cite{Habib:1990hx,LaFlamme:1990kd,Nambu:1991vs,Prokopec:1992ia,Polarski:1995jg}, but has largely been used for exploring the decoherence of fluctuations in cosmological backgrounds~\cite{Grishchuk:1990bj,Burgess:2006jn,Kiefer:2008ku,Burgess:2014eoa,Boyanovsky:2015tba,Burgess:2022nwu,Burgess:2024eng}. It is a practical fact that cosmological correlators are effectively classical, but the manner in which classicality emerges remains an area of active research. A potentially audacious theoretical and observational goal is to determine empirically that structure in the universe arises from quantum fluctuations~\cite{Maldacena:2015bha,Martin:2015qta,Green:2020whw,Espinosa-Portales:2022yok,Piotrak:2025zhy}. Such a determination would necessitate techniques beyond the Bell inequalities~\cite{Martin:2015qta,dePutter:2019xxv} and instead would have to isolate the quantum evolution during the inflationary epoch. A more precise understanding of the evolution of the density matrix in cosmology may serve as a setting stone to these ambitions as well.

\paragraph{Acknowledgments}
We are grateful to Prish Chakraborty, Tim Cohen, Jordan Cotler, Tarun Grover, Yiwen Huang, Yue-Zhou Li, Juan Maldacena, John McGreevy, Enrico Pajer, Geoff Penington, Akhil Premkumar, Zipei Zhang and Guanhao Sun. DG and KG are supported by the US~Department of Energy under grant~\mbox{DE-SC0009919}. 

\newpage
\appendix


\section{Relationship Between Generating Functionals and Probabilities}\label{app:connections}

In the main text, we chose to derive the density matrix of the local long distance field $\varphi$ by defining the generating functional
\beq
Z[J_\pm,t] = {\rm Tr} \left[e^{J_+ \bar\varphi} e^{J_- \bar \pi} \rho_{\rm red}(\varphi_+(\x),\varphi_-(\x))  \right] \ .
\eeq
While working with the generating function is a common framework, it does make the connection to previous work less transparent. The purpose of this section is to show this representation is identical to the previous approaches. To make the discussion easy to follow, we will work with a collection of harmonic oscillator wavefunctions $\psi_i(x)$ but we want to calculate a generating functional for $\bar x = \sum_k \hat x_k$ and $\bar p = N^{-1} \sum_k \hat p_k$, namely
\beq
Z[J_\pm] = {\rm Tr}_k \left[ e^{J_+ \bar x} e^{J_- \bar p}  \prod_k |\psi_k \rangle \langle \psi_k | \right] ={\rm Tr} \left[ e^{J_+ \bar x} e^{J_- \bar p} \rho \right] \ ,
\eeq
where $\rho = \prod_k \rho_k$ is the total density matrix for all $k$-harmonic oscillators.

First, we would like to compare this with our random walks discussed in Section~\ref{sec:walks}, where we defined a density matrix for a degree of freedom $\hat X$ via the interaction
\beq
\rho(X,X') = {\rm Tr}_{k} \left[ e^{i \hat P_X \bar x } |X=0\rangle \langle X'= 0 |   \, e^{-i \hat P_{X'} \bar x} \otimes \rho \right]  \ .
\eeq
where we took the trace in the $\hat x_k$ basis. Now we notice that if we work in the momentum basis $X\to P_X$, $|X =0\rangle  \to {\rm constant}$ and therefore
\beq
\rho(P_+,P_{-}) = {\rm Tr}_{k} \left[ e^{i  P_+ \bar x } \rho e^{-i  P_{-} \bar x} \right] =  Z[J_+ = i (P_+- P_-),J_- = 0]  \ ,
\eeq
where $P_{\pm}$ are just the momentum eigenvalues. We see that in the momentum basis, the density matrix is just the generating function with imaginary values of $J$. In fact, this also represents the measurements of a local observer along a worldline as in~\cite{}. 

Another approach is to calculate the density matrix while imposing the constraint $\sum x_i =  \bar x$ in the form of a $\delta$-function,
\beq
\rho(\bar x+\Delta x, \bar x -\Delta x) = {\rm Tr}_k \left[ \delta(\bar x+\Delta x -\sum_k x_k) \delta(\bar x-\Delta x  -\sum_k x'_k) \rho(\{ x_i\},\{x'_i\}) \right] \ .
\eeq
This was employed in \cite{Gorbenko:2019rza} for $\Delta x =0$ in order to determine the probability distribution in stochastic inflation for $\bar \varphi \to \bar x$. In order to derive such a result from the generating function, we first note that if we take $J_- = i 2\Delta x$, then 
\beq
Z[J_+,i \Delta x] = {\rm Tr}[e^{J_+ \bar x}\rho(\{x+\Delta x/N \},\{ x'-\Delta x/N\}) ]
\eeq 
Now defining $J_+ = i k$ and taking the Fourier transform we get 
\bea
&&\int \frac{dk}{(2\pi)} e^{-i k \bar x} Z[J_+=i k, J_- = i \Delta x \ .] \\
&&= {\rm Tr}_k \left[ \delta(\bar x -\sum_k x_k) \rho(\{ x_i+ \Delta x/N \},\{x_i-\Delta x/N\}) \right] \ .
\eea
Applying this formula back to the de Sitter context, if we take $\Delta x=0$, $\bar x \to \bar \varphi$, and $x_k \to \varphi(\k)$ we get
\beq
P(\bar \varphi) = \int D \varphi \, \delta \left(\bar \varphi - \int \frac{d^3k}{(2\pi)^3} \varphi(\k) \right) \,  P[\varphi(\k)]
\eeq
in agreement with \cite{Gorbenko:2019rza}. Finally, from this expression, we can derive the Wigner quasi-probability distribution for $\bar x$ and $\bar x'$. Starting from,  
\bea
W[\bar x, \bar p] = \int d \Delta x e^{-i p \Delta x} \rho \left(\bar x + \frac{\Delta x}{2},  \bar x - \frac{\Delta x}{2} \right) \ .
\eea
We can transform $Z[J_\pm]$ as the following
\begin{equation}
    \begin{aligned}
    \Tr_k \bigg[e^{J_+ \bar{x}} e^{J_- \bar p} \rho \bigg] = \Tr_k \bigg[  e^{\frac{iJ_x J_p}{2}}e^{J_x \bar{x}} e^{\frac{J_-  \bar{p}}{2}} \rho e^{\frac{J_-  \bar{p}}{2} } \bigg] = \Tr_k \bigg[e^{\frac{iJ_x J_p}{2}} e^{J_x \bar x} \rho\left(x - \frac{iJ_-}{2N}, x' + \frac{i J_-}{2N} \right) \bigg]
    \end{aligned}
\end{equation}
Then, taking $J_+ = ik$, $J_- = iy$, we get $W[\bar{x}, \bar{p}]$ via
\begin{equation}
    W[\bar x, \bar{p}] = \int dy \,e^{-i\bar p y} \int \frac{dk}{(2\pi)} e^{i\frac{ky}{2}} e^{-ik \bar{x}} Z[J_+ = ik, J_-=iy] 
\end{equation}
This, of course, is not limited to the Wigner distribution alone; any Fourier transform of the Wigner distribution can be derived in the same way. 

\section{Evolution equation for the generating functional}\label{app:generating_func_evolution}
In this section, we want to simplify the evolution equation for the generating functional. We start with the definition
\beq
Z[J_\pm,t] = {\rm Tr} \left[ e^{J_+ \bar\varphi} e^{ J_- \bar \pi} \rho_{\rm red}(\varphi_+(\x),\varphi_-(\x))  \right] \ ,
\eeq
and hence we have
\bea
\frac{d}{dt}  Z[J_\pm,t] &= {\rm Tr} \left[e^{J_+ \bar\varphi}e^{J_- \bar \pi} \frac{d}{dt} \rho_{\rm red}(\varphi_+(\x),\varphi_-(\x))  \right]  \ .
\eea 
At this stage, we can substitute the evolution equation for $\rho_{\rm red}$  using \eqref{eq:Master_Lindblad_eq}. Defining ${\cal O} = e^{J_+ \bar\varphi} e^{J_- \bar \pi}$, we have that 
\begin{equation}
    {\cal O} (\varphi_-(p), \varphi_+(p)) =  \bra{\varphi_-} e^{J_+ \bar\varphi} e^{J_- \bar \pi} \ket{\varphi_+} = e^{J_+ \bar \varphi_-} \delta \left[\varphi_+(p) - \varphi_- (p)+ \frac{i J_- K(p^2/\Lambda^2)}{N_k} \right] \ .
\end{equation}
Then, we equivalently have 
\begin{equation}
\begin{aligned}
    Z[J_\pm, t] &= \int D \varphi_+ D \varphi_- {\cal O}(\varphi_-, \varphi_+) \rho(\varphi_+, \varphi_-)\\
    &= \int D\varphi_+\, D\varphi_- e^{J_+ \bar \varphi_-} \delta \left[\varphi_+(p) - \varphi_- (p) + \frac{i J_- K(p^2/\Lambda^2)}{N_k} \right] \rho(\varphi_+, \varphi_-) \ .
\end{aligned}
\end{equation}
Now, we can plug in our time derivative of $\rho$ to get

\begin{align}
        \frac{d Z}{dt}&= \int D \varphi_+ \, D \varphi_- {\cal O} \int \frac{d^3p}{(2\pi)^3}\, \,  \bigg[\frac{H\Delta_0}{2}  \left( \frac{\delta}{\delta \varphi_+} + \frac{\delta}{\delta \varphi_-} \right)^2 - a^6(t_0)  K^{-2} \frac{H\Delta_2}{2} (\varphi_- -\varphi_+)^2  \\
        & + i a^3(t_0)K^{-1}  \frac{H\Delta_R}{2} \bigg\{(\varphi_+-\varphi_-), \left(\frac{\delta}{\delta \varphi_+} + \frac{\delta}{\delta \varphi_-} \right) \bigg\} + 2a^3 K^{-1} H\Delta_{I} \left(  \frac{\delta}{\delta \varphi_{+}}  + \frac{\delta}{\delta \varphi_{-}} \right)(\varphi_{+} + \varphi_{-})  \bigg] \nonumber \\
        &+\frac{i K}{2a^3} \left(\frac{\delta^2}{\delta \varphi_+^2} - \frac{\delta^2}{\delta \varphi_-^2} \right)\rho - \frac{i K^{-1} a^3}{2}\left (\frac{(\nabla \varphi_+)^2}{a^2} - \frac{(\nabla \varphi_-)^2}{a^2}  \right )\rho - i a^3(V(\varphi_+) - V(\varphi_-)) \rho \ . \nonumber
\end{align}
We can integrate by parts until we simplify each term. All the derivatives will ultimately be put onto ${\cal O}$. The rules simplify to 
\begin{equation} \label{eq:rules_homogenization}
\begin{gathered}
    {\cal O}\left( \frac{\delta}{\delta \varphi_+} + \frac{\delta}{\delta \varphi_-}\right) \rho \to -J_+ {\cal O} \rho \qquad {\cal O}(\varphi_+ - \varphi_-) \rho \to -\frac{i J_- K}{N_k} {\cal O} \rho \\
    {\cal O}\left( \frac{\delta}{\delta \varphi_+} - \frac{\delta}{\delta \varphi_-}\right) \rho \to \left(2\frac{N_k}{iK} \frac{\partial}{\partial J_-} \right) {\cal O} \rho \ .
\end{gathered}
\end{equation}
We will also need to make a further simplification. In order to deal with the $\varphi_+ + \varphi_-$ terms, we will resort to perturbation theory and use the following leading order identity (for the massless case) as explained in Appendix~\ref{app:homogenization}
\begin{equation}
    \begin{aligned}
        (\varphi_+ + \varphi_-)\rho &= -K\frac{H^2}{k^3} \left(\frac{\delta}{\delta \varphi_+} + \frac{\delta}{\delta \varphi_-} \right)\rho + \frac{i}{k \eta}(\varphi_+ - \varphi_-) \rho\\
        {\cal O} (\varphi_+ + \varphi_-) \rho &\to K\frac{H^2}{k^3}J_+ {\cal O \rho} - aH\frac{J_-K}{N_k k} {\cal O} \rho
    \end{aligned}
\end{equation}
Using these identities, we can completely simplify the above equation. We get 

\begin{align}
    \frac{dZ}{d t} &= \frac{1}{2} \int \frac{d^3p}{(2\pi)^3} \bigg[ H\Delta_0 J_+^2 + a^6 H\Delta_2 \left ( \frac{J_- }{N_k} \right )^{2} - 2 a^3 H \Delta_R \left ( \frac{ J_- }{N_k} J_+ \right ) \\
	&- 2a^3 \Delta_I  \frac{H^{3}}{p^3} J_+^2 + 2a^4 \frac{H^2}{p} \Delta_I \left ( \frac{J_+ J_-}{N_k} \right ) \bigg] K' Z + \frac{N_{k}}{a^{3}} J_+ \left( -\frac{i J_{+}}{2} + \frac{\partial}{\partial J_{-}} \right) Z  \nonumber \\
    &- ia^3 \left(\tilde{V}\left(-iJ_+ - i \frac{J_- }{2N_k} \right) - \tilde{V}\left(-iJ_+ + i \frac{J_- }{2N_k} \right) \right) Z  + \int \frac{d^3p}{(2\pi)^3} \left ( K\frac{a H^2}{2p} \frac{J_+ J_-}{N_k} - \frac{a^2H p}{2} K \frac{J_-^2}{N_k^2} \right)Z \nonumber
 \end{align}
The terms with the $p$ integral correspond to the shell of momentum just integrated in, while the local terms correspond to Hamiltonian time evolution. After using the form of the $\Delta$ for the massless theory, we get 
\begin{equation}
    \begin{aligned}
        \frac{dZ[J_\pm]}{dt} &= \frac{1}{2} \int \frac{d^{3} p}{(2 \pi)^{3}}\bigg[ \left(\frac{H^{3}}{2p^{3}} J_+^2 - a^2(t_0)H \frac{p}{2} \frac{(J_-)^2}{N_k^2} - \frac{aH^2}{p} \frac{J_-J_+}{N_k} \right) \frac{\partial K}{\partial s} \bigg] Z[J_\pm]\\
        & +\frac{N_{k}}{a^{3}} J_+\left( -\frac{i J_{+}}{2} + \frac{\partial}{\partial J_{-}} \right) Z[J_\pm] - \frac{J_{-}}{N_{k}} a^3 \tilde{V}' \left(-iJ_+ \right) Z[J_\pm] \\
        &+\int \frac{d^3p}{(2\pi)^3} \left ( K\frac{a H^2}{2p} \frac{J_+ J_-}{N_k} - \frac{a^2H p}{2} K \frac{J_-^2}{N_k^2} \right)Z[J_\pm]\\
    \end{aligned}
\end{equation}

\subsection{Identity useful for homogenization}\label{app:homogenization}
The homogenization of terms of the form 
\begin{equation}
    \int \frac{d^3p}{(2\pi)^3} {\cal O}(\varphi_-, \varphi_+) (\varphi_+ + \varphi_-)(p) \rho_R
\end{equation}
need to be dealt with perturbatively. The basic idea is that for harmonic oscillators, this represents a term of the form $x \psi \propto p \psi$ when $\psi$ is a ground state. We can use a similar idea here, starting with the UV mode $\phi$, that is defined for the free theory as
$$
\phi(k) = \frac{H}{\sqrt{ 2k^{3}}} [(1 - ik \eta) e^{ ik \eta } a_{k} + (1 + ik \eta )e^{ -ik \eta }a_{k}^{\dagger} ] \qquad \Pi(k) = K^{-1}\frac{\sqrt{ k }}{\sqrt{ 2 } H \eta}[ e^{ ik \eta } a_{k} + e^{ -ik \eta } a_{k}^{\dagger} ]  \ .
$$
Solving for $a_k$ in the limit $k \eta \ll 1$, we get
\begin{equation}
 \begin{aligned} 
a_{k} &= i [\phi(k, \eta) v_{k}^\star - K\Pi(k, \eta) u_{k}^\star], \qquad  v_{k} = a^{2} u'_{k} \ ,\qquad u_{k}= \frac{H}{\sqrt{ 2k^{3} }}(1 + ik \eta)e^{ -ik \eta } \ ,\\
&= i \left[ \frac{k^{2}}{H \sqrt{ 2k^{3} }\eta}(1 + ik \eta) \phi(k, \eta) - \frac{H}{\sqrt{ 2k^{3} }} K\Pi(k, \eta)  \right] \\
&= \frac{i}{\sqrt{ 2k^{3} }} \left[  \left( \frac{k^{2}}{H \eta} + \frac{ik^{3}}{H} \right) \phi - H K \Pi\right] \ .
 \end{aligned} 
\end{equation}
The lowering operator annihilates the vacuum, giving 
$$
a_{k}\Psi_{\rm BD} = 0 \implies \left( \frac{k^{2}}{H \eta} + \frac{ik^{3}}{H}  \right) \phi(k) \Psi_{\rm BD} = -iH K\frac{\delta}{\delta \phi(k)} \Psi_{\rm BD} \ ,
$$
as can be checked directly using
$$
\Psi_{BD}[\phi, \eta]
= \prod_{\vec{k}} \left( \frac{2 k^3}{\pi} \right)^{1/4}
\exp\!\left[
\frac{i K^{-1}}{2 H^2} \frac{k^2}{\eta (1 - i k \eta)} \,
\phi(\vec{k}) \phi(-\vec{k})
\right]
\frac{e^{-i k \eta / 2}}{\sqrt{1 - i k \eta}} \, .
$$
Thus, the density matrix satisfies
\begin{equation}
 \begin{aligned} 
H^{2}K\left( \frac{\delta}{\delta \phi_{+}} + \frac{\delta}{\delta \phi_{-}} \right) \rho = \left( \frac{ik^{2}}{\eta}(\phi_{+} - \phi_{-}) - k^{3}(\phi_{+} + \phi_{-})\right) \rho \ .
 \end{aligned} 
\end{equation}
This is a useful simplification. We will write this as 
\begin{equation}
    (\phi_+ + \phi_-)\rho = -\frac{H^2}{k^3} K\left(\frac{\delta}{\delta \phi_+} + \frac{\delta}{\delta \phi_-} \right)\rho + \frac{i}{k \eta}(\phi_+ - \phi_-) \rho \ .
\end{equation}
This is, of course, only true at the Gaussian level, which is sufficient for our purposes right now. Now, we can evaluate terms of the form 
\begin{equation}
    \int \frac{d^3p}{(2\pi)^3} {\cal O}(\varphi_-, \varphi_+) (\varphi_+ + \varphi_-)(p) \rho_R \to \int  \frac{d^3p}{(2\pi)^3} {\cal O} \left( -\frac{H^2}{p^3} K\left(\frac{\delta}{\delta \varphi_+} + \frac{\delta}{\delta \varphi_-} \right)\rho + \frac{i}{p \eta}(\varphi_+ - \varphi_-) \rho_R \right) 
\end{equation}
using the rules in \eqref{eq:rules_homogenization}.

\section{Eliminating Momentum with Moments}\label{app:moments}

The type of random walk under consideration is generally referred to as a ``Fast-Slow" system in the random walk literature. The fast variables have evolution with a time scale $\gamma^{-1}$ where the slow system evolves on a times scale set by the potential $V'$ and the amplitude of the noise, $A$. One then aims to systematically eliminate the fast variables, leaving a Markovian evolution equation for the slow variables as an expansion in $\gamma^{-1}$. There is a formal procedure~\cite{gardiner2004handbook} where the solutions are projected onto the solutions of the fast equations, giving a systematic expansion in $\gamma^{-1}$. However, as we are only interested in calculating the NLO (and potentially NNLO) terms, this procedure is both overkill and somewhat ill-suited for the problem at hand. 

The most straightforward approach to eliminating $Q$ from the evolution of $W(\bv,Q)$, following~\cite{Li:2025azq}, is to define the moments of the distribution
\beq
W_N(\bv) = \int dQ (a^{-3}Q)^N W(\bv,Q)  \qquad P(\bf) \equiv W_0(\bf) \ .
\eeq
We have changed variables such that the evolution equation is schematically of the form
\begin{align}
    \frac{d}{dt}W[\bar \varphi, Q] &=  \frac{H^{3}}{8\pi ^{2}} \frac{\partial^2 W}{\partial \varphi ^{2}} +  a^{3} V_{\mathrm{eff}}'' (\bv)\frac{ H^{2}}{12 \pi ^{2}}  \frac{\partial^2 W}{\partial Q \partial \bv }  + a^6 \frac{H}{72 \pi ^{2}} (V_{\mathrm{eff}}''(\bv))^{2} \frac{\partial ^{2} W}{\partial Q^{2}} \nonumber\\
    & -  V_{\mathrm{eff}}''(\bv) \frac{\partial}{\partial Q} \left(\frac{Q}{3H} W \right) - \frac{Q}{a^{3}} \frac{\partial W}{\partial \bv} + \frac{1}{3H}  \frac{\partial}{\partial \bv} (V_{\mathrm{eff}}'(\bv) W) \ ,
\end{align}
In terms of the moments, this equation becomes
\bea\label{eq:moments}
\left(\frac{d}{dt} + 3H N\right) W_N[\bar \varphi]&=&  \hat L_1 W_N - \frac{\partial W_{N+1}}{\partial \bar \varphi} \nonumber \\
    && +  V''_{\rm eff} \left( -N \frac{H^2}{12\pi^2}  \frac{\partial}{\partial \bar \varphi}W_{N-1} + \frac{H}{72 \pi^2}  V''_{\rm eff} N(N-1) \partial W_{N-2}\right) \ ,
\eea
where $\hat L_1 W= \frac{\partial}{\partial \bar \varphi} \left(\frac{1}{3H} V_{\rm eff}'(\bar \varphi)\right) W  + \frac{H^3}{8 \pi^2} \frac{\partial^2 W}{\partial \bar \varphi^2} + N \frac{V''_{\rm eff}}{3H} W$. Therefore, we have
\bea
\frac{d}{dt} W_0 &=&\hat L_1 W_0 - \frac{\partial}{\partial \bv} W_1 \\
\frac{d}{dt} W_1 + 3H W_1 &=& L_1 W_1 - V''_{\rm eff} \frac{H^2}{12 \pi^2}\frac{\partial}{\partial \bv} W_0 + \ldots
\eea
Here the $\ldots$ are higher order in $\lambda$ and can be dropped for the NLO terms. At this order, it's easy to solve for $W_1$ treating $3H = \gamma$ as the large parameter, so that 
\beq
\frac{d}{dt} W_0 =\hat L_1 W_0 + \frac{1}{3H}\frac{\partial}{\partial \bv} V''_{\rm eff} \frac{H^2}{12 \pi^2}\frac{\partial}{\partial \bv} W_0 \ .
\eeq
We see therefore both the correction to the effective potential and the noise at NLO.

At higher orders, such as NNLO, the natural question is whether it is eliminating the higher orders in $W_N$ or non-linear contributions to $S_{\rm eff}$ that will play the dominant role in determining the evolution equations. We will be specifically interested in whether the following NNLO term can explain the results of~\cite{Cohen:2021fzf},
\beq
\frac{d}{dt}\tilde W(\bv,Q)\supset - \frac{1}{24} V'''(\varphi)a^9 \frac{\partial^3}{\partial Q^3} W \ .
\eeq
This gives rise to a new contribution to $W_3$,
\beq
\left(\frac{d}{dt} + 9H\right) W_3(\bar \varphi) \supset \frac{1}{4} V'''(\bv) W_0(\bv) \ .
\eeq
Using Equation~(\ref{eq:moments}), we can solve for leading order in $H^{-1}$ terms to solve for the contributions to $W_2$ and then $W_1$, to find
\bea
W_2 \approx -\frac{1}{6 H} \frac{\partial}{\partial \bv} W_3  &=& -\frac{1}{216 H^2}\frac{\partial}{\partial \bv}V'''(\bv) W_0(\bv) \\
W_1 \supset -\frac{1}{3 H} \frac{\partial}{\partial \bv} W_2  &=& \frac{1}{648 H^3}\frac{\partial^2}{\partial \bv^2}V'''(\bv) W_0(\bv) \ , 
\eea
and therefore our correction to the evolution of $W_0\equiv P$ becomes
\beq
\frac{d}{dt} P  \supset -\frac{1}{648 H^3} \frac{\partial^3}{\partial \bv^3}\left( V'''(\bv) P \right) \ .
\eeq
This is of the correct form to describe out non-Gaussian noise term at NNLO but the coefficient does not appear to match the result from SdSET~\cite{Cohen:2021fzf}. Here we have neglected many other NNLO terms, which are generally not universal, along with any corrections to $S_{\rm eff}$ in the interacting theory. This is therefore an incomplete calculation and shows the need for more efficient techniques for calculating at NNLO and beyond.

\clearpage
\phantomsection
\addcontentsline{toc}{section}{References}
\small
\bibliographystyle{utphys}
\bibliography{Refs}

\end{document}